\documentclass[journal,draftcls,onecolumn,12pt,a4paper]{IEEEtran}
\usepackage{amssymb}
\usepackage{amsfonts}
\usepackage{epstopdf}
\usepackage{graphicx}
\usepackage{stmaryrd}
\usepackage{amssymb,amsmath}
\usepackage{multirow,url}
\usepackage{color,cite}
\usepackage[OT1]{fontenc} 
\usepackage{comment}
\usepackage{subcaption}

\ifodd 1
\newcommand{\com}[1]{\textbf{\color{red} (COMMENT: #1)}} 
\newcommand{\comg}[1]{\textbf{\color{green} (COMMENT: #1)}}
\newcommand{\response}[1]{\textbf{\color{magenta} (RESPONSE: #1)}} 
\else

\newcommand{\com}[1]{}
\newcommand{\comg}[1]{}
\newcommand{\response}[1]{}
\fi
\ifodd 0
\newcommand{\referred}[1]{\textcolor{red}{RefPaper: #1}} 
\else
\newcommand{\referred}[1]{}
\fi

\ifodd 0
\newcommand{\changeblue}[1]{\textcolor{blue}{Modified: #1}} 
\else
\newcommand{\changeblue}[1]{}
\fi

\begin{document}

\title{Age of Information in Physical-Layer Network Coding Enabled Two-Way Relay Networks}

\author{Haoyuan~Pan,~\IEEEmembership{Member,~IEEE,}~Tse-Tin~Chan,~\IEEEmembership{Member,~IEEE,}\\~Victor~C.~M.~Leung,~\IEEEmembership{Life Fellow,~IEEE,}~Jianqiang~Li%

\thanks{H. Pan, Victor C. M. Leung, and J. Li are with the College of Computer Science and Software Engineering, Shenzhen University, Shenzhen, 518060, China (e-mails: {hypan@szu.edu.cn}, {vleung@ieee.org}, {lijq@szu.edu.cn}). }
\thanks{T.-T.~Chan is with the Department of Computing, The Hang Seng University of Hong Kong, Hong Kong (e-mail: {ttchan@hsu.edu.hk}).}
}

\maketitle
\thispagestyle{empty}

\begin{abstract}
This paper investigates the information freshness of two-way relay networks (TWRN) operated with physical-layer network coding (PNC). Information freshness is quantified by age of information (AoI), defined as the time elapsed since the generation time of the latest received information update. PNC reduces communication latency of TWRNs by turning superimposed electromagnetic waves into network-coded messages so that end users can send update packets to each other via the relay more frequently. Although sending update packets more frequently is potential to reduce AoI, how to deal with packet corruption has not been well investigated. Specifically, if old packets are corrupted in any hop of a TWRN, one needs to decide the old packets to be dropped or to be retransmitted, e.g., new packets have recent information, but may require more time to be delivered. We study the average AoI with and without automatic repeat request (ARQ) in PNC-enabled TWRNs. We first consider a non-ARQ scheme where old packets are always dropped when corrupted, referred to as once-lost-then-drop (OLTD), and a classical ARQ scheme with no packet lost, referred to as reliable packet transmission (RPT). Interestingly, our analysis shows that neither the non-ARQ scheme nor the pure ARQ scheme achieves good average AoI. We then put forth an uplink-lost-then-drop (ULTD) protocol that combines packet drop and ARQ. Experiments on software-defined radio indicate that ULTD significantly outperforms OLTD and RPT in terms of average AoI. Although this paper focuses on TWRNs, we believe the insight of ULTD applies generally to other two-hop networks. Our insight is that to achieve high information freshness, when packets are corrupted in the first hop, new packets should be generated and sent (i.e., old packets are discarded); when packets are corrupted in the second hop, old packets should be retransmitted until successful reception.
\end{abstract}

\begin{IEEEkeywords}
Age of information, information freshness, physical-layer network coding, ARQ
\end{IEEEkeywords}


\IEEEpeerreviewmaketitle



\section{Introduction}\label{sec:intro}
In recent years, age of information (AoI) has been regarded as a key performance metric to measure information freshness in the next-generation communication networks \cite{aoi_mono,AoIMagazine,AoI}. In many real-time monitoring and control systems, such as traffic monitoring in vehicular networks \cite{aoi_first} and motion control in industrial Internet of Things (IIoT) \cite{IIOT}, timely delivery of regular and frequent information updates is crucial since correct decision making or precise control rely on real-time data. Prior studies show that conventional performance metrics, such as information rate and packet delay, are not appropriate to quantify information freshness in the emerging timely information update systems \cite{aoi_mono,AoIMagazine}. 

AoI is defined as the time elapsed since the generation time of the latest received information update at the destination \cite{aoi_mono,AoIMagazine,AoI}. Specifically, AoI captures both the delay and the generation time of each information update $-$ if at time $t$, the latest information update received by the receiver was an update packet generated at the source at time $t'$, then the instantaneous AoI at time $t$ is $t-t'$. Since AoI characterizes the effect of packet delay from the perspective of the destination, it is quite different from the conventional delay/latency metric. For example, most prior works studied the ``average AoI'', defined as the time average of instantaneous AoI over a long period \cite{AoIMagazine}. The analysis and optimization of average AoI in various networks showed that age optimality conditions often do not coincide with those of conventional metrics such as throughput and latency \cite{aoi_mono}. 

\begin{figure}
\centering
\includegraphics[width=0.6\textwidth]{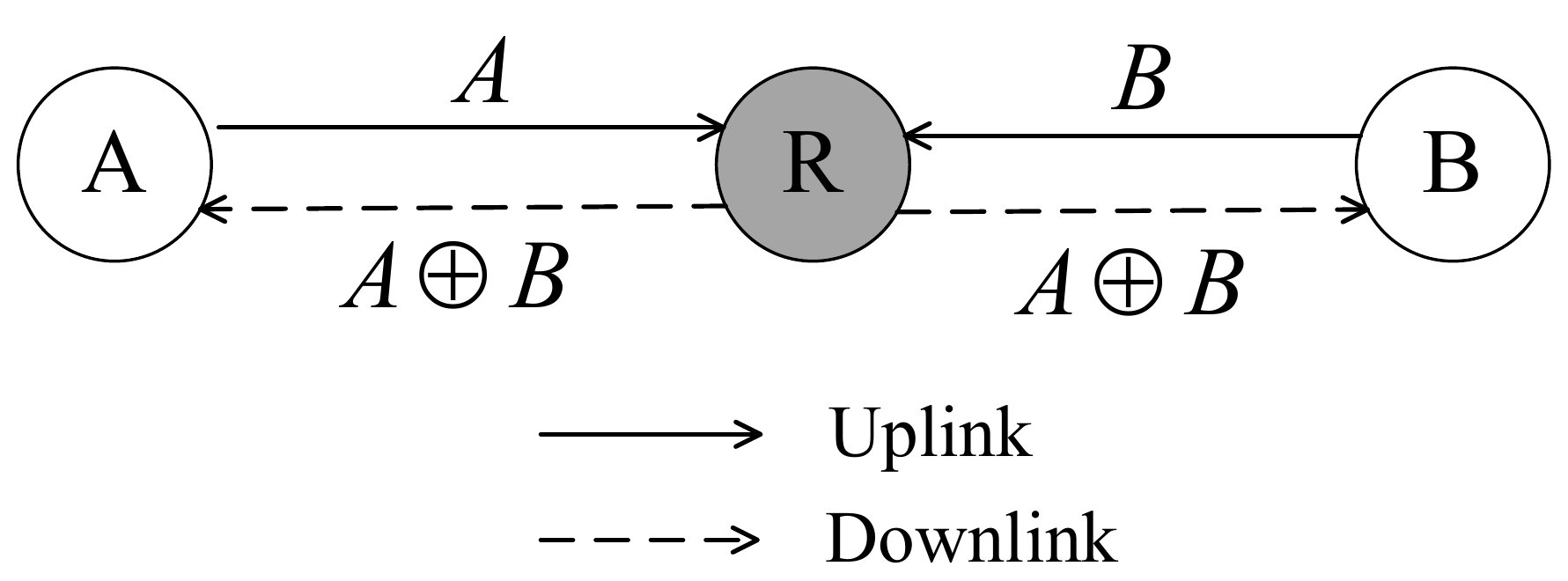}
\caption{A two-way relay information update system, where two end users A and B want to send update packets to each other with the help of a relay.}
\label{fig:system_model}
\end{figure}

In a common set-up of information update systems, two end users send their latest update packets to each other with the help of a relay over a wireless medium, as shown in Fig. \ref{fig:system_model}. Such a network is referred to as a two-way relay network (TWRN) in the literature \cite{PNC06}. TWRN-type information update systems can be found in many scenarios. For example, in vehicular networks, two vehicles send periodic basic safety messages (BSM) to each other via a road side unit (RSU) as a relay \cite{hybridmac}. A BSM generated by each vehicle usually contains a vehicle's instant status information, such as velocity, direction, acceleration, and position. Since a periodic exchange of BSMs creates mutual awareness of the surrounding environment, it is crucial to receive fresh BSMs to reduce the risk of road accidents.

The traditional store-and-forward relaying scheme requires four time slots in total for the two end users to deliver a packet to each other (e.g., two time slots for each user to send a packet to the other user) \cite{PNC06}. Physical-layer network coding (PNC) is a key technique to reduce communication latency and to improve network throughput \cite{liew2015primer,PNC06}. It turns mutual interference between wireless signals from simultaneously transmitting users into useful network-coded information. To see this, let us focus on a PNC-enabled TWRN shown in Fig. \ref{fig:system_model}. Compared with the traditional store-and-forward relaying, PNC reduces the total number of time slots to two for the exchange of two packets. Specifically, the first time slot is an uplink phase in which the two end users send packets simultaneously to the relay. In the second time slot, the relay performs PNC decoding on the superimposed received signals and broadcasts back a network-coded packet to the end users in the downlink phase \cite{liew2015primer,PNC06}. We refer to the network-coded packet as a PNC packet, an eXclusive-OR (XOR) of the two source packets from the end users. Upon receiving the PNC packet, the two end users subtract their self-packet from the PNC packet to obtain the packet from the other user. Therefore, PNC has halved the transmission time and improved the throughput twice of a TWRN \cite{liew2015primer,PNC06}. 

In the context of information update systems, this paper considers the generate-at-will model where the source can sample the latest information of the observed phenomenon at any given time \cite{AoIMagazine}. Since PNC reduces the time for a user to receive packets from the other user, end users can sample and send update packets to each other more frequently. When packet transmission is successful in every hop, a more frequent update leads to higher information freshness (i.e., lower AoI). In practice, however, update packets are typically short. Information theory reveals that with finite block lengths, packet error rates (PER) cannot go to zero \cite{gallager}. Although PNC is promising to achieve high information freshness, it has not been well investigated when update packets get corrupted in the uplink or the downlink phase. In particular, how to deal with the corrupted packets becomes a critical issue to achieve low average AoI, especially in a TWRN with two hops. This paper is an attempt to fill this gap.

Traditional wireless communication systems use automatic repeat request (ARQ) to ensure reliable transmission by packet retransmission \cite{tse2005fundamentals}. That is, if a packet corrupts, the receiver sends a negative acknowledgement (NACK) to the transmitter via a feedback channel, and then the corrupted packet (i.e., the same packet) is retransmitted until it is successfully received. Otherwise, an acknowledgement (ACK) is sent back to the transmitter, and a new packet is sent. However, applying ARQ to information update systems directly may lead to high average AoI since the number of retransmissions could be arbitrarily large. Prior studies on a single-hop network indicate that a non-ARQ scheme where a new packet is sent immediately once an old packet got corrupted outperforms the classical ARQ scheme significantly \cite{aoi_arq}. This is because new packets always have the latest information (old packets become obsolete once new packets are generated). However, in a two-hop TWRN, new packets may require more time to be delivered (e.g., when packet corruption occurs in the second hop). Therefore, a quantitative investigation is needed to study the relative merits between ARQ and packet drop. 

In this paper, we first investigate the average AoI of a PNC-enabled TWRN under two protocols, namely once-lost-then-drop (OLTD) and reliable packet transmission (RPT). In OLTD, once a packet gets corrupted in either the uplink or the downlink phase, the two end users immediately generate and send new packets to the relay, i.e., a non-ARQ scheme without any packet retransmission. By contrast, in RPT, a link-by-link ARQ is used in both the uplink and the downlink transmission to ensure that there is no packet lost. 

Interestingly, unlike single-hop networks, our analysis shows that OLTD only improves little over RTP, indicating that neither a non-ARQ protocol (e.g., OLTD, where old packets are always dropped) nor a pure ARQ protocol (e.g., RPT, where old packets are always retransmitted) can achieve low average AoI in TWRNs. The problem of OLTD is that the PNC packet decoded at the relay is sent only once in the downlink, regardless of the decoding result. If the downlink packet corrupts, a retransmission could have helped recover it (and hence a successful update instantly), but OLTD blindly goes back to the uplink phase, leading to a waste of time. To reduce the average AoI, we put forth an uplink-lost-then-drop (ULTD) protocol that combines the advantages of both OLTD and RPT. More specifically, in the uplink phase, when the relay fails to decode the PNC packet, old packets are dropped and the two end users transmit new packets to the relay immediately. In the downlink phase, ARQ is used to ensure reliable transmission. In other words, ULTD combines both packet drop and ARQ to improve information freshness of TWRNs.

For performance evaluation, we compare the average AoI performance of different protocols theoretically and experimentally. We first use Gallager's random coding bound (RCB) to estimate the PER of short packets under different block lengths (e.g., the PERs of both uplink and downlink in a TWRN). We find the optimal block length that leads to the minimum average AoI, and compare the average AoI of the three protocols. For concept proving in a practical setting, we conduct real experiments on software-defined radio. Both theoretical and experimental results show that ULTD significantly outperforms RPT and OLTD in terms of average AoI. For example, when the SNR is as low as 7.5dB, ULTD reduces the average AoI by 25\% and 32\% compared with OLTD and RPT, respectively. Overall, thanks to the combination of ARQ and packet drop, ULTD is a preferable protocol to achieve high information freshness in TWRN operated with PNC.

To sum up, we have three major contributions:
\begin{itemize}\leftmargin=0in
\item [(1)] We study a PNC-enabled TWRN with AoI requirements. Specifically, we are the first to investigate the ARQ protocols in TWRNs to deal with the corrupted packets in both uplink and downlink transmissions, aiming to achieve high information freshness.
\item [(2)] We design an uplink-lost-then-drop (ULTD) protocol for PNC-enabled TWRNs. In particular, ULTD combines both packet drop (to drop old packets and to send new packets in the uplink) and ARQ (to retransmit old packets in the downlink until successful reception), a key principle to achieve low AoI in two-hop TWRNs.
\item [(3)] We demonstrate the practical feasibility of ULTD on software-defined radio. Our experiments show that ULTD outperforms the classical ARQ scheme and the non-ARQ scheme in terms of average AoI in a practical PNC-enabled TWRN setting.
\end{itemize}

\section{Related Work}\label{sec:relatedwork}
Age of information (AoI), as a new performance metric, has attracted considerable research interests in recent years \cite{aoi_mono,AoIMagazine}. It was first proposed to characterize the timeliness of safety packets in vehicular networks \cite{aoi_first}. Later, AoI has been studied under different communication and network models. Prior AoI works focused more on the upper layers of the communication protocol stack (i.e., above the PHY and MAC layers), and we refer the readers to the monograph \cite{AoI} and the references therein for the important research results. For example, early works focused on analyzing the AoI performance of different systems modeled by various abstract queueing models, in which information update packets arrive at the source node randomly following a memoryless Poisson process \cite{AoI,AoI_single_server,queue1,AoIqueue}. Different scheduling policies \cite{AoIscheduling4,AoIscheduling3,AoIscheduling2,AoIscheduling1} are then examined with the aim of minimizing different AoI metrics, such as average AoI \cite{AoI}, peak AoI \cite{peak_aoi}, bounded AoI \cite{aoi_tdma_fdma}, etc. By contrast, the current work does not study queueing models: there is no queue at the source nodes. We adopt a generate-at-will model, wherein a source node will make a measurement and generate an update packet only when it has the chance to transmit \cite{AoIMagazine}. 

Moving down to the PHY and MAC layers, prior works optimized the average AoI in various wireless scenarios with packet retransmission to deal with wireless impairments, including both automatic request control (ARQ) and hybrid ARQ (HARQ) \cite{aoi_arq,aoi_harq,chenhe_aoi,steam_code,aoi_multicast}. In particular, recent studies have shown that packet preemption provides significant advantages in terms of average AoI, e.g., to drop and to replace old packets with new packets, since new packets always contain more up-to-date information \cite{control_drop}. For example, \cite{chenhe_aoi} proposed a truncated ARQ scheme where the same old update packet was sent only finite times until the maximum allowable transmission times. Moreover, advanced coding schemes are proposed to combine old packets and new packets into the same transmission as a form of ARQ or HARQ. Specifically, when an old packet is retransmitted, a new packet is embedded in the old packet using different methods, e.g., joint packet coding is used in \cite{aoi_adaptive_coding, steam_code}. In particular, in order to improve information freshness, dropping old packets properly is shown to be an effective way to balance the error corrections of old packets and fast decodings of new packets in \cite{steam_code}. Motivated by these previous studies, this paper considers both packet drop and ARQ to lower the average AoI of a two-way relay network (TWRN).

Despite the above efforts, only a few AoI studies have been dedicated to networks beyond one hop. In particular, the impacts of packet drop and ARQ schemes in each hop on AoI have not been well investigated. Ref. \cite{aoi_multihop} abstracted a large multi-hop network as a directed graph and developed scheduling policies to achieve age-optimality or near age-optimality with a single information flow; \cite{BoostingorHindering} studied AoI and throughput optimization in routing-aware multi-hop networks. Still, they do not focus on PHY or MAC layers. Refs. \cite{mangang_relay1,mangang_relay2} considered the age and energy tradeoff in a one-way multi-relay network with two hops. They modeled the PHY-layer PER by short-packet theory and studied the relay selection strategy. However, no packet drop or ARQ is studied in \cite{mangang_relay1,mangang_relay2}. In addition, unlike the single information flow in \cite{aoi_multihop,mangang_relay1,mangang_relay2}, the TWRN considered in this paper contains two information flows, e.g., two end users exchange update packets with the help of a relay.

Very recently, \cite{aoi_noma_tii,aoi_af} investigated the potentials of applying PNC to information update systems. PNC was first proposed in \cite{PNC06} to turn the mutual interference (superposed signals) of multiple transmitters into useful network-coded information at a receiver. PNC has been studied and evaluated in depth: we refer the interested readers to \cite{liew2015primer} for details. For the ﬁrst time, \cite{aoi_noma_tii} applied PNC to reduce the average AoI of a non-orthogonal multiple access (NOMA) network. The authors show that by combining PNC with multiuser decoding techniques, the decoded PNC packets significantly reduce the average AoI compared with conventional orthogonal multiple access (OMA). However, the results of \cite{aoi_noma_tii} cannot be applied to the current work directly because \cite{aoi_noma_tii} studied a single-hop network without relays. Ref. \cite{aoi_af} evaluated the average AoI of TWRNs where the relay adopts an amplify-and-forward strategy. The received superimposed signals are amplified at the relay and forward to end users (e.g., a special case of PNC \cite{liew2015primer}). By contrast, our work adopts a decode-and-forward strategy where the relay tries to decode PNC packets. In particular, we study how to deal with the corrupted packets to improve information freshness. 

\section{Preliminaries}\label{sec:preliminaries}
\subsection{System Model}\label{sec:preliminaries1}
We consider a two-way relay network (TWRN) where two end users A and B want to send information update packets to each other with the help of a relay, as shown in Fig. \ref{fig:system_model}. In information update systems, each end user (say, user A) wants to receive the update packets of the other end user (say, user B) as timely as possible. 

\begin{figure*}
\centering
\includegraphics[width=1\textwidth]{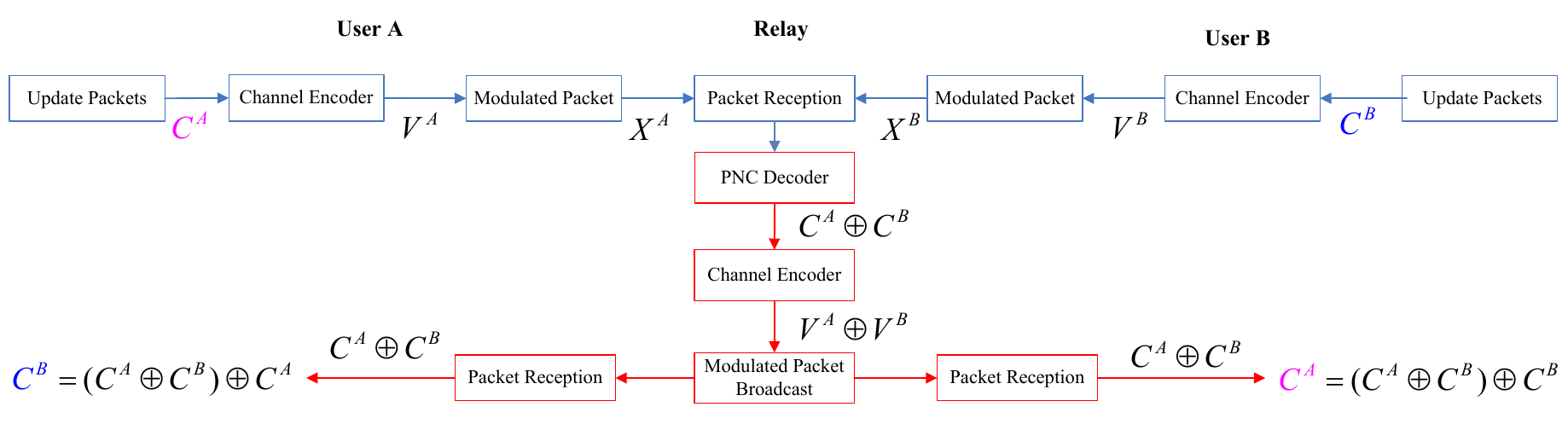}
\caption{The general architecture of information processing at the end nodes and the relay in a PNC-enabled TWRN.}
\label{fig:general_architecture}
\end{figure*}

In a TWRN operated with PNC, only two time slots are required for the two end users to communicate with each other via a relay. Specifically, as shown in Fig. \ref{fig:system_model}, the first time slot is an uplink transmission phase from both end users to the relay, and the second time slot is a downlink transmission phase from the relay to the end users. In time slot 1, as illustrated in Fig. 
\ref{fig:general_architecture}, the two users generate their respective packets ${C^A}$ and ${C^B}$, which are then channel-coded into ${V^A}$ and ${V^B}$, respectively. When the two users transmit simultaneously to the relay, the relay receives superimposed signals. A PNC decoder attempts to decode a linear combination of two packets ${C^A}$ and ${C^B}$ from the superimposed signals, i.e., a PNC packet $C^A \oplus C^B$ that is the eXclusive-OR (XOR) of ${C^A}$ and ${C^B}$. In time slot 2, the relay broadcasts $C^A \oplus C^B$ to both end users. An end user can recover the packet of the other user with the help of the received PNC packet $C^A \oplus C^B$ and its own packet, e.g., for user A, packet ${C^B}$ can be obtained by $C^B = (C^A \oplus C^B) \oplus C^A$. 

Throughout this paper, we adopt the XOR-channel decoding (XOR-CD) approach for PNC decoding in the uplink \cite{liew2015primer}. XOR-CD works because it exploits the linearity of linear channel codes such as convolutional codes. Specifically, if we define $\Gamma ( \cdot )$ as the convolutional encoding operation, we have 
\begin{align}
\Gamma \left( {C^A \oplus C^B} \right) = \Gamma \left( {C^A} \right) \oplus \left( {C^B} \right) = V^A \oplus V^B.
\end{align}
As such, in the XOR-CD decoder, it first passes the received superimposed signals through a PNC demodulator to obtain bit-wise XOR information, i.e., the XOR bits $V^A \oplus V^B$, and then feeds these XOR bits to a standard Viterbi decoder (as used in a conventional 802.11 WLAN system) to decode the PNC packet $C^A \oplus C^B$. The XOR bits can be soft bits or hard bits. This paper considers soft bit-wise XOR information, and the details of the computation of the soft bits can be found in Appendix \ref{sec:PNC} and \cite{liew2015primer}.

Compared with the traditional non-network-coded relaying scheme, PNC reduces the number of transmission time slots from four to two for the exchange of two packets between user A and user B \cite{liew2015primer}. Although PNC is well-known to reduce the communication latency of TWRNs, its merits have not been well investigated when the system performance metric is the information freshness. In the following, we introduce the metric to evaluate information freshness in this paper, namely age of information (AoI).

\subsection{Age of Information (AoI)}\label{sec:preliminaries2}
Suppose that the $i$-th update packet from user $j,j \in \{ A,B\}$ is generated at time instant $t_i^j$ and is received by the other user at time instant ${\tilde t_i}^j$. At any time instant $t$, the packet index of the last received update from user $j$ is ${N_j}(t) = \max \{ i|{\tilde t_i}^j < t\} $. Hence, the generation time of the most recently received update from user $j$ is ${G_j}(t) = {t_{{N_j}(t)}}$. The instantaneous AoI of user A, ${\Delta _A}(t)$, measured at user B, is  
\begin{align}
{\Delta _A}(t) = t - {G_A}(t).
\end{align}
\noindent Correspondingly, the instantaneous AoI of user B, ${\Delta _B}(t)$, measured at user A, is ${\Delta _B}(t) = t - {G_B}(t)$. The instantaneous AoI ${\Delta _j}(t)$ is a continuous-time continuous-value stochastic process \cite{aoi_mono}. A smaller instantaneous AoI means higher information freshness. 

With the instantaneous AoI ${\Delta _j}(t)$, we can compute other AoI metrics. This paper uses \emph{average AoI}, a widely studied AoI metric, to evaluate the information freshness of TWRNs \cite{aoi_mono}. Average AoI is defined as the time average of the instantaneous AoI. Specifically, the average AoI of user $j$ is given by 
\begin{align}
{\bar \Delta _j} = \mathop {\lim }\limits_{T \to \infty } \frac{1}{T}\int_0^T {{\Delta _j}(t)} dt.
\end{align}
\noindent A lower average AoI indicates that the update packets of user $j$ are generally fresher over a long period. 

\subsection{Random Coding Bound}\label{sec:preliminaries3}
This paper uses the random coding bound (RCB) to estimate the packet error rate (PER) in the TWRN. In practical information update systems with AoI requirements, update packets are typically short. Information theory tells us that with finite block lengths, the PER cannot go to zero \cite{gallager}. Suppose that in the uplink, the relay can successfully decode the superimposed signals to a PNC packet with probability $\alpha $. In the downlink, we assume that the two end users have an equal probability $\beta $ to decode the broadcast PNC packet. In general, for a coded packet, $\alpha $ and $\beta $  increase as the coded block length increases. 

We first consider the downlink phase. The downlink transmission can be regarded as two point-to-point channels. For point-to-point channels, Gallager's RCB on the average block error probability $1 - \beta$ of random $(L, K)$ codes has the form \cite{gallager}
\begin{align}
1 - \beta  \le {2^{ - L{E_G}(R)}},
\label{equ:rcb}
\end{align}
\noindent where $K$ is the number of source bits of a packet; $L$ is the block length of coded bits; $R=K/L$ is the code rate. ${E_G}(R)$ is the random coding error exponent \cite{gallager}. Under perfect channel state information at the receiver, ${E_G}(R)$ is 
\begin{align}
{E_G}(R) = {\max _{0 \le \rho  \le 1}}[{E_0}(\rho ) - \rho R],
\label{equ:rcb_eg}
\end{align}
\noindent where $\rho$ is the auxiliary variable over which optimization is performed to get the maximum value of the right-hand side of (\ref{equ:rcb_eg}), and
\begin{align}
{E_0}(\rho ): =  - {\log _2}E\left[ {{{\left( {\frac{{E[{p_{Y|C}}{{(Y|C')}^{\frac{1}{{1 + \rho }}}}|Y]}}{{{p_{Y|C}}{{(Y|C)}^{\frac{1}{{1 + \rho }}}}}}} \right)}^\rho }} \right],
\label{equ:rcb_e0}
\end{align}
where $C$  and $C'$ are the two independent and uniformly distributed binary random variables, e.g., the coded bits having values 0 or 1. $Y$ is the received signal is an AWGN channel. 

For the TWRN uplink transmission, \cite{short_pnc} shows that the RCB (\ref{equ:rcb}) holds also for the ensemble of linear random codes and thus applies to PNC systems. In particular, (\ref{equ:rcb}) applies to PNC decoding that employs an XOR-CD decoder at the receiver. Specifically, the PER performance of the XOR-CD decoder can be fully characterized by analyzing the transmission with a linear block code over a ``virtual'' memory-less point-to-point channel \cite{short_pnc}. The difference of XOR-CD is that $C$ and $C'$ in  (\ref{equ:rcb_e0}) are now the two independent and uniformly distributed binary random variables associated with the XOR-coded bits (rather than the individual coded bits in point-to-point channels). 

With non-zero PERs, ARQ protocols are commonly used to ensure reliable transmission in wireless communication systems. However, when the system performance metric is information freshness, this paper asks a question: in order to achieve low average AoI, should the corrupted packets be retransmitted, especially in a TWRN network with two hops? Since new packets always have the latest information, Section \ref{sec:OLTD} studies the average AoI of a PNC-enabled TWRN without ARQ. Specifically, when old packets are corrupted in the uplink or the downlink phase, new packets are generated and transmitted (i.e., old packets are dropped). We refer to this scheme as the once-lost-then-drop (OLTD) protocol. Later in Section \ref{sec:ARQ}, we evaluate the average AoI with ARQ. Notice that ARQ can be used to ensure a reliable uplink or downlink transmission, or even both. The average AoI under different strategies requires a thorough investigation.

\section{Average AoI in PNC-enabled TWRNs without ARQ}\label{sec:OLTD}
This section analyzes the average AoI in PNC-enabled TWRNs without ARQ, where the system adopts a so-called once-lost-then-drop (OLTD) protocol. In OLTD, once old packets get corrupted either in the uplink phase (i.e., the relay fails to decode a PNC packet from the superimposed signals from the two end users) or in the downlink phase (i.e., an end user fails to decode the broadcast PNC packet), the two end users generate and transmit new packets to the relay immediately. In other words, old packets are always dropped and there is no packet retransmission in OLTD. 

\begin{figure}
\centering
\includegraphics[width=0.6\textwidth]{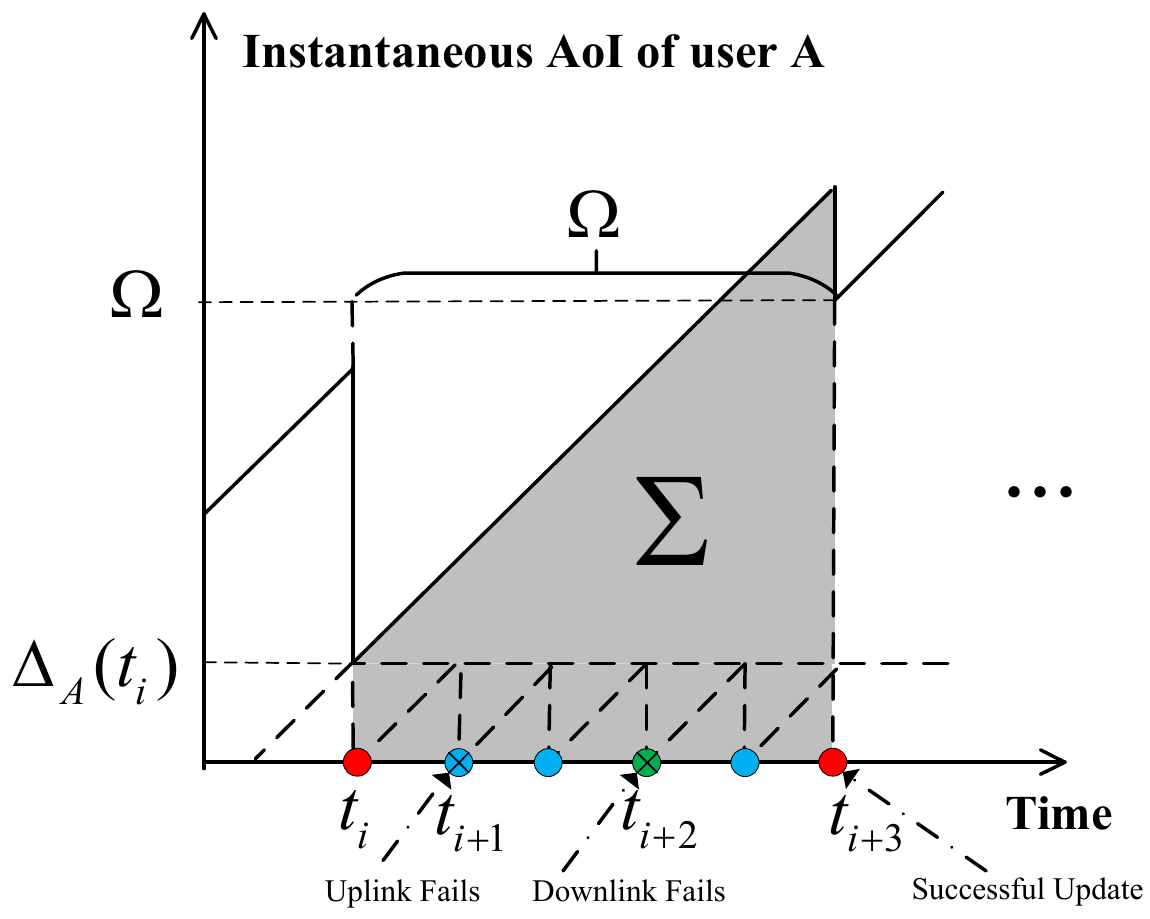}
\caption{An example of user A's instantaneous AoI, ${\Delta _A}$, under the once-lost-then-drop (OLTD) protocol. In round $i$, users A and B send packets $C_i^A$ and $C_i^B$ to the relay simultaneously. Since there is no ARQ, round $i$ ends when the relay fails to decode a PNC packet $C_i^A \oplus C_i^B$ in the uplink. New packets $C_{i + 1}^A$ and $C_{i + 1}^B$ are sent in round $i+1$, but user A cannot receive the downlink PNC packet. An update is successful in round $i+2$, and ${\Delta _A}$ drops to $\Omega$, which is the total duration from round $i$ to round $i+2$.}
\label{fig:aoi_oltd}
\end{figure}

\subsection{Once-Lost-Then-Prop (OLTD)  Protocol Description}\label{sec:OLTD1}
We consider a time-slotted system where each time slot is an uplink or a downlink transmission in TWRNs. The packet duration in the uplink or the downlink occupies one time slot. Notice that with OLTD, uplink and downlink transmissions may not occur in alternative time slots due to packet corruption. The detailed operations of the OLTD protocol are as follows:
\begin{itemize}\leftmargin=0in
\item \textbf{In the uplink phase}, the two users send their update packets simultaneously to the relay, and the relay tries to decode a PNC packet from the superimposed signals of the two uplink packets. If an error occurs, the relay sends a NACK to the end users via a feedback channel. The two end users then generate and transmit two new packets to the relay immediately in the next time slot. Otherwise, the decoded PNC packet is broadcast to the end users in the next time slot (i.e., a downlink phase).
\item \textbf{In the downlink phase}, the two users try to decode the broadcast PNC packet. Whether the downlink PNC packet is decoded or not, both users generate and send new update packets to the relay in the next time slot. In other words, the PNC packet is broadcast only once regardless of the decoding outcome at the end users.
\end{itemize}

Fig. \ref{fig:aoi_oltd} plots an example of user A's instantaneous AoI ${\Delta _A}$, measured at user B. In Fig. \ref{fig:aoi_oltd}, the evolution of  ${\Delta _A}$ between two consecutive successful updates is depicted. After a successful update, denote by $\Omega$ as the time needed for the next update to be successfully received at user B. That is, in OLTD, $\Omega$ is the time required for user B to receive a packet from user A, starting from the time instance of the last successful update. 

We define a round as the duration of the time interval between end users sending two new update packets. In round $i$, denote by $C_i^A$ and $C_i^B$ the two update packets sent by user A and user B, respectively. As shown in Fig. \ref{fig:aoi_oltd}, packets $C_i^A$ and $C_i^B$ are sent at time ${t_i}$. Suppose that the relay fails to decode a PNC packet $C_i^A \oplus C_i^B$ in round $i$ (i.e., an uplink failure). Since there is no packet retransmission in OLTD, round $i$ ends and new packets $C_{i + 1}^A$ and $C_{i + 1}^B$ are sent in round $i+1$. Now the relay decodes the PNC packet $C_{i + 1}^A \oplus C_{i + 1}^B$ that is broadcast to the end users. Here we assume that user A cannot receive $C_{i + 1}^A \oplus C_{i + 1}^B$ (i.e., a downlink failure). Again, there is no packet retransmission in the downlink, so round $i+2$ starts and new packets $C_{i + 2}^A$ and $C_{i + 2}^B$ are sent at  ${t_{i + 2}}$. Assuming both uplink and downlink transmissions are successful in round $i+2$, the instantaneous AoI of user A ${\Delta _A}$ drops to two time slots at the end of round $i+2$. Also, $\Omega$ is the total duration from round $i$ to round $i+2$ in Fig. \ref{fig:aoi_oltd}. 

\subsection{Average AoI in OLTD}\label{sec:OLTD2}
We now analyze the average AoI under OLTD using the renewal process theory. The whole time interval $(0,T)$ contains a series of consecutive renewal periods (e.g., successful information updates). For simplification, we normalize the time slot duration to one in the following analysis. To compute the average AoI $\bar \Delta _j^{OLTD}$ of OLTD, $j \in \{ A,B\}$, let us consider the area $\Sigma$ between the two consecutive successful updates at ${t_i}$  and  ${t_{i+3}}$ (i.e., a renewal period), as shown in Fig. \ref{fig:aoi_oltd}. The area of $\Sigma$ is computed by
\begin{align}
\Sigma = {\Delta _j}({t_i})\Omega  + \frac{1}{2}{\Omega ^2} = 2\Omega  + \frac{1}{2}{\Omega ^2},
\end{align}
\noindent where ${\Delta _j}({t_i}) = 2$ since a successful update always drops the instantaneous AoI to two time slots in OLTD. The average AoI $\bar \Delta _j^{OLTD}$ is 
\begin{align}
\bar \Delta _j^{OLTD} &= \mathop {\lim }\limits_{T \to \infty } \frac{1}{T}\int_0^T {{\Delta _j}(t)} dt = \mathop {\lim }\limits_{W \to \infty } \frac{{\sum\nolimits_{w = 1}^W {{A_{(w)}}} }}{{\sum\nolimits_{w = 1}^W {{\Omega _{(w)}}} }}  \notag \\
&= \frac{{E\left[ {2\Omega  + \frac{1}{2}{\Omega ^2}} \right]}}{{E\left[ \Omega  \right]}} = 2 + \frac{{E\left[ {{\Omega ^2}} \right]}}{{2E\left[ \Omega  \right]}}.
\end{align}

\begin{figure}
\centering
\includegraphics[width=0.6\textwidth]{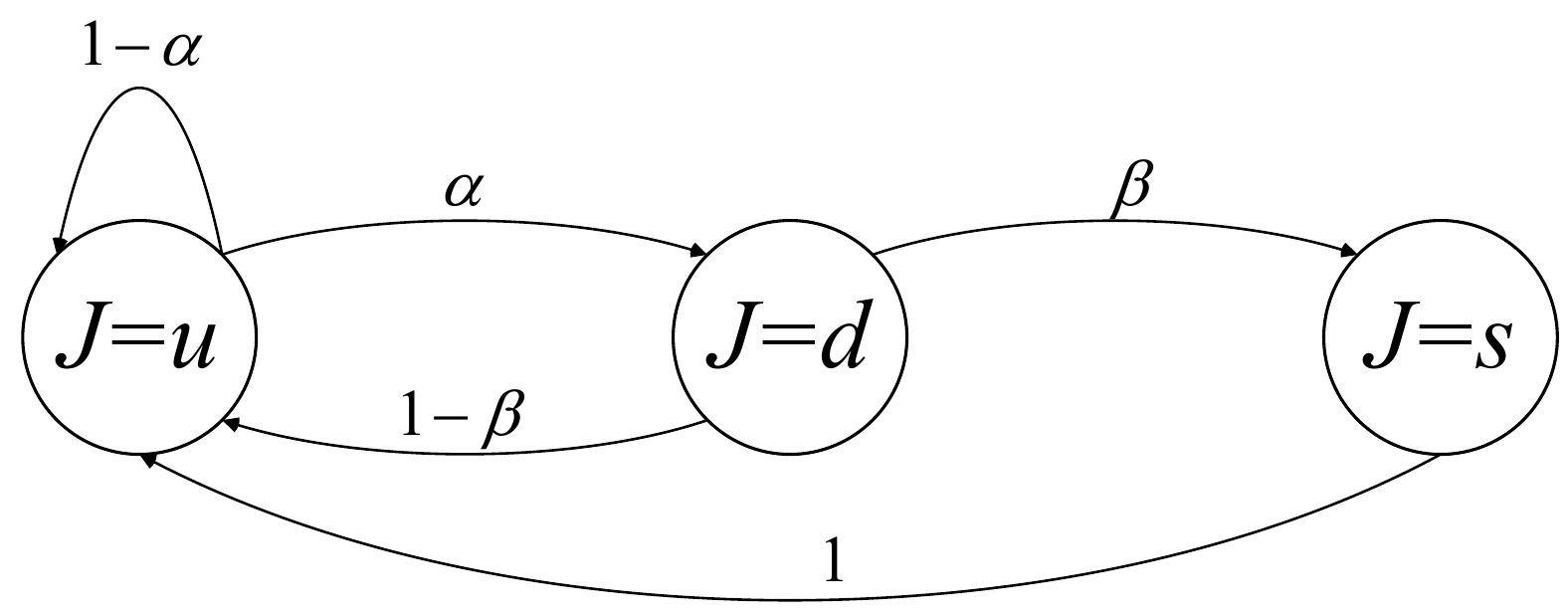}
\caption{The Markov model for the OLTD protocol. State $J = \{ u,d,s\}$ represents the stages of the packet transmission in a TWRN: $u$ stands for $u$plink; $d$ stands for $d$ownlink; $s$ stands for $s$uccessful reception at the destination.}
\label{fig:oltd_markov}
\end{figure}

In order to compute $\bar \Delta _j^{OLTD}$, we use a Markov model depicted in Fig. \ref{fig:oltd_markov} to compute $E[\Omega ]$ and $E[{\Omega ^2}]$. In the Markov model, the state $J = \{ u,d,s\}$ represents the stages of the packet transmission in a TWRN. $J = u$ represents an uplink transmission from the end users to the relay; $J = d$ represents a downlink transmission from the relay to the end users; $J = s$ represents a successful update of user $j$ (e.g., if $j$=A, it means that user B successfully receives an update packet from user A). More specifically, at state $J = u$, the system transits to $J = d$ if the relay decodes a PNC packet with probability $\alpha$, and remains in $J = u$ if the relay fails to decode a PNC packet with probability $1-\alpha$ (i.e., to drop the old packets). At state $J = d$, the system transits to $J = s$ if the broadcast PNC packet is decoded with probability $\beta$ (e.g., user $j$’s update packet can be recovered); otherwise, with probability $1-\beta$, the system transits to an uplink transmission (i.e., $J = u$; to drop the old packets). Finally, at state $J = s$, the system will transit to $J = u$ for a new round with probability one.

Let $\Pi $ denote the state transition matrix, which can be written as 
\begin{align}
\Pi  = \left( {\begin{array}{*{20}{c}}
{{\pi _{uu}}}&{{\pi _{ud}}}&{{\pi _{us}}}\\
{{\pi _{du}}}&{{\pi _{dd}}}&{{\pi _{ds}}}\\
{{\pi _{su}}}&{{\pi _{sd}}}&{{\pi _{ss}}}
\end{array}} \right) = \left( {\begin{array}{*{20}{c}}
{1 - \alpha }&\alpha &0\\
{1 - \beta }&0 &\beta\\
1&0&0
\end{array}} \right),
\end{align}
\noindent where ${\pi _{mn}}$ is the probability of transiting from state $J = m$ to state $J = n$, for $m,n \in \{ u,d,s\} $. Since $\Omega$ is the time needed for the next update to be successfully received,  $\Omega$ equals the time going through states from state $J = u$ to state $J = s$ in the Markov model, e.g., the state transition goes through states ${J_0} = u,{J_1},{J_2},...,{J_\Omega } = s$. Denote by ${\tau _{mn}}$ the expected time required to transverse from state $J = m$ to state $J = n$. Then, ${\tau _{mn}}$ can be expressed as
\begin{align}
{\tau _{mn}} = E[{t_n}|{J_0} = m],
\end{align}
\noindent where ${t_n}$ is a random variable that represents the time to reach state $J = n$ for the first time. By definition, $E[\Omega ]$ equals ${\tau _{us}}$, and we compute ${\tau _{us}}$ by
\begin{align}
{\tau _{us}} &= E[{t_s}|{J_0} = u]\notag \\
&= 1 + E[{t_s}|{J_1} = u]\Pr ({J_1} = u|{J_0} = u) + E[{t_s}|{J_1} = d]\Pr ({J_1} = d|{J_0} = u) \notag\\
&= 1 + {\tau _{us}}{\pi _{uu}} + {\tau _{ds}}{\pi _{ud}} \notag\\
 &= 1 + (1 - \alpha ){\tau _{us}} + \alpha {\tau _{ds}}.
\label{equ:tau_us}
\end{align}

In (\ref{equ:tau_us}), ${\tau _{ds}}$ is computed by
\begin{align}
{\tau _{ds}} &= E[{t_s}|{J_0} = d]\notag \\
& = 1 + E[{t_s}|{J_1} = u]\Pr ({J_1} = u|{J_0} = d) + E[{t_s}|{J_1} = s]\Pr ({J_1} = s|{J_0} = d)\notag \\
&=1 + {\tau _{us}}{\pi _{du}}\notag \\
& = 1 + (1 - \beta ){\tau _{us}}.
\label{equ:tau_ds}
\end{align}
Substituting (\ref{equ:tau_ds}) into (\ref{equ:tau_us}), ${\tau _{us}}$ and $E[\Omega ]$ are computed and simplified as  
\begin{align}
E[\Omega ] = {\tau _{us}} = \frac{{1 + \alpha }}{{\alpha \beta }}.
\label{equ:omega_exp1}
\end{align}

Similarly, to compute $E[{\Omega ^2}]$, we define ${\lambda _{mn}}$ as the expectation of the second moment of the time required to transverse from state $J = m$ to state $J = n$ for the first time. Then, ${\lambda _{mn}}$ can be expressed as
\begin{align}
{\lambda _{mn}} = E[{\left( {{t_n}} \right)^2}|{J_0} = m].
\label{equ:omega_exp1}
\end{align}
By definition, $E[{\Omega ^2}]$ equals ${\lambda _{us}}$, which is computed by
\begin{align}
{\lambda _{us}} &= E[{\left( {{t_s}} \right)^2}|{J_0} = u]\notag \\
& = E[{\left( {1 + {t_s}} \right)^2}|{J_0} = u]\Pr ({J_1} = u|{J_0} = u) + E[{\left( {1 + {t_s}} \right)^2}|{J_1} = d]\Pr ({J_1} = d|{J_0} = u)\notag \\
& = 1 + 2\left( {{\tau _{us}}{\pi _{uu}} + {\tau _{ds}}{\pi _{ud}}} \right) + \left( {{\lambda _{us}}{\pi _{uu}} + {\lambda _{ds}}{\pi _{ud}}} \right)
\label{equ:lambda_us}
\end{align}

\noindent where ${\lambda _{ds}}$ is 
\begin{align}
{\lambda _{ds}} &= E[{\left( {{t_s}} \right)^2}|{J_0} = d]\notag \\
&=E[{\left( {1 + {t_s}} \right)^2}|{J_1} = u]\Pr ({J_1} = u|{J_0} = d) + E[{\left( {1 + {t_s}} \right)^2}|{J_1} = s]\Pr ({J_1} = s|{J_0} = d)\notag \\
& = 1 + 2{\tau _{us}}{\pi _{du}} + {\lambda _{us}}{\pi _{du}}
\label{equ:lambda_ds}
\end{align}

Substituting (\ref{equ:lambda_ds}) into (\ref{equ:lambda_us}),  ${\lambda _{us}}$ and $E[{\Omega ^2}]$ are now computed and simplified as
\begin{align}
E[{\Omega ^2}] = {\lambda _{us}} = \frac{{2 + 4\alpha  + 2{\alpha ^2} - 3\alpha \beta  - {\alpha ^2}\beta }}{{{\alpha ^2}{\beta ^2}}}.
\label{equ:omega_exp2}
\end{align}

Finally, with $E[{\Omega}]$ (\ref{equ:omega_exp1}) and $E[{\Omega ^2}]$ (\ref{equ:omega_exp2}), we obtain the average AoI $\bar \Delta _j^{OLTD}$
\begin{align}
\bar \Delta _j^{OLTD} = 2 + \frac{{E\left[ {{\Omega ^2}} \right]}}{{2E\left[ \Omega  \right]}} = 2 + \frac{{2 + 4\alpha  + 2{\alpha ^2} - 3\alpha \beta  - {\alpha ^2}\beta }}{{2(1 + \alpha )\alpha \beta }}.
\label{equ:aoi_oltd_final}
\end{align}

In (\ref{equ:aoi_oltd_final}), the first term 2 means the instantaneous AoI once an update packet is received (i.e., two time slots is the lowest instantaneous AoI in a TWRN with two hops). When both $\alpha$ and $\beta$ equal to one, $\bar \Delta _j^{OLTD}$ has the minimum average AoI of 3 time slots. However, when $\alpha$ and $\beta$ are not one, we evaluate $\bar \Delta _j^{OLTD}$ by RCB and experiments, as will be presented in Section \ref{sec:ARQ3} and Section \ref{sec:exp}, respectively.

\section{Average AoI in PNC-enabled TWRNs with ARQ}\label{sec:ARQ}
This section presents the average AoI in TWRNs with ARQ. Specifically, we first consider reliable packet transmission where ARQ is used in both the uplink and the downlink phase. After that, we consider ARQ is only used in the downlink phase and the old packets are dropped if they are lost in the uplink phase. A theoretical comparison among different schemes using RCB is presented in Section \ref{sec:ARQ3}.

\subsection{Reliable Packet Transmission (RPT)}\label{sec:ARQ1}
We first study the average AoI of a TWRN assuming reliable packet transmission (RPT). By ``reliable'', we mean that a link-by-link ARQ is used so that retransmission is issued when the uplink or the downlink packet corrupts (i.e., there is no packet lost in either the uplink or the downlink phase). The detailed operations of the RPT protocol are as follows:
\begin{itemize}\leftmargin=0in
\item \textbf{In the uplink phase}, the two users send their update packets simultaneously to the relay, and the relay tries to decode a PNC packet. If an error occurs, the relay sends a NACK to the end users via a feedback channel. The same packets are retransmitted until the PNC packet is successfully decoded at the relay. When the PNC packet is decoded, it is broadcast to the end users in the downlink phase.
\item \textbf{In the downlink phase}, the two users try to decode the downlink PNC packets. If an error occurs, a user sends a NACK to the relay via a feedback channel. The PNC packet is then retransmitted until successful reception at both end users. We remark that an end user may successfully receive the PNC packet before the other end user. The relay has to continue sending the PNC packet when only one user receives it. When both users receive the PNC packet, they sample and send new update packets to the relay in the next time slot.
\end{itemize}

\begin{figure}
\centering
\includegraphics[width=0.6\textwidth]{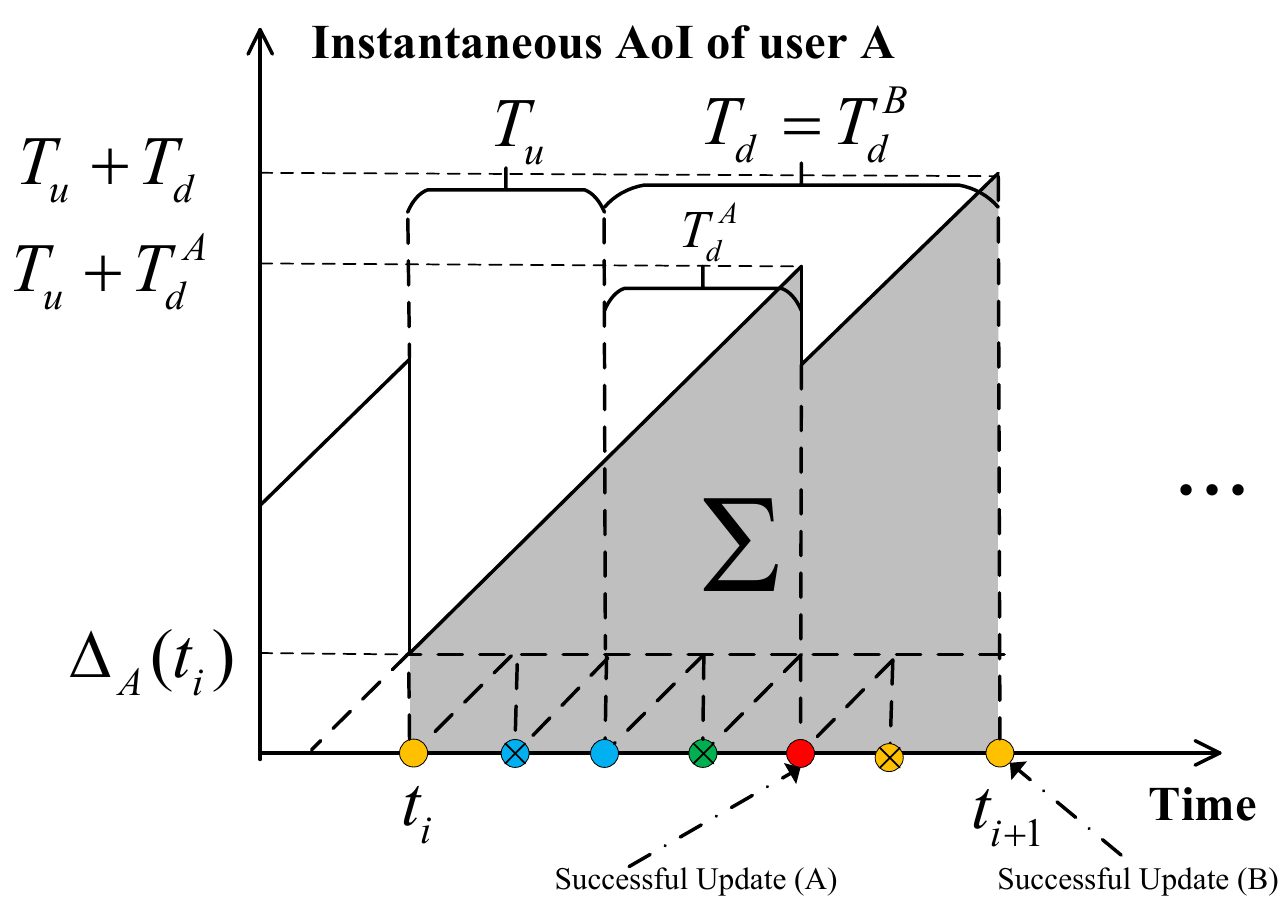}
\caption{An example of user A's instantaneous AoI, ${\Delta _A}$, with reliable packet transmission (RPT) in round $i$. In the uplink transmission, the two users transmit the same packets $C_i^A$ and $C_i^B$ twice so that the relay decodes a PNC packet $C_i^A \oplus C_i^B$ at  ${t_i} + {T_u}$. In the downlink transmission, user B decodes $C_i^A \oplus C_i^B$ first so that user B recovers user A's packet first (at time ${t_i} + {T_u} + T_d^A$), so ${\Delta _A}$ is reset to ${T_u} + T_d^A$. ${\Delta _A}$ continues to increase until the end of round $i$ when user B's packet is recovered at  ${t_{i + 1}}={T_u} + T_d^B$.}
\label{fig:aoi_rpt}
\end{figure}

As in OLTD, we define a round as the duration between end users sending two new update packets. With RPT, a round is the duration between the time when two new packets are sent for the first time in the uplink and the time when both users receive the PNC packet in the downlink. To see this, Fig. \ref{fig:aoi_rpt} plots an example of the evolution of user A's instantaneous AoI ${\Delta _A}$ with RPT. We see that there is only one round between the two consecutive successful updates.

As shown in Fig. \ref{fig:aoi_rpt}, suppose that round $i$ starts at time ${t_i}$ and ends at time ${t_{i + 1}}$. In the uplink transmission, the relay cannot decode a PNC packet when $C_i^A$ and $C_i^B$ are sent for the first time. Later, retransmission of $C_i^A$ and $C_i^B$ is issued, and the relay decodes the PNC packet $C_i^A \oplus C_i^B$ successfully at time ${t_i} + {T_u}$, where ${T_u}$ is the total duration of the uplink phase. In the downlink, the relay broadcasts $C_i^A \oplus C_i^B$ four times in total until both users receive it. Specifically, user B decodes $C_i^A \oplus C_i^B$ when it is sent at the second time. Using the received $C_i^A \oplus C_i^B$, $C_i^A$ is recovered at user B and the instantaneous AoI ${\Delta _A}$ drops to ${T_u} + T_d^A$, where $T_d^A$ is the time required for $C_i^A$ to be recovered in the downlink phase. We also use $T_d^B$ to denote the time required for $C_i^B$ to be recovered at user A. In this example, user A recovers $C_i^B$ when $C_i^A \oplus C_i^B$ is sent at the fourth time. Since $C_i^B$ is recovered later than  $C_i^A$, after ${\Delta _A}$ is reset to ${T_u} + T_d^A$, ${\Delta _A}$ continues to increase linearly until the end of the round. We use ${T_d}$ to denote the total duration of the downlink phase, e.g., ${T_d} = T_d^B$ since $T_d^A<T_d^B$ in Fig. \ref{fig:aoi_rpt}.

\textbf{Average AoI in RPT}: We now analyze the average AoI in RPT $\bar \Delta _j^{RPT}$, $j \in \{ A,B\}$. Similar to OLTD, we first consider the area $\Sigma$ shown in Fig. \ref{fig:aoi_rpt}. The area of $\Sigma$ is computed by 
\begin{align}
\Sigma &= {\Delta _j}({t_i})({T_u} + T_d^j) + \frac{1}{2}{({T_u} + T_d^j)^2} + ({T_u} + T_d^j)({T_d} - T_d^j) + \frac{1}{2}{({T_d} - T_d^j)^2}\notag \\
& = ({T_u}' + {T_d}')({T_u} + T_d^j) + ({T_u} + T_d^j)({T_d} - T_d^j) + \frac{1}{2}{({T_u} + T_d^j)^2} + \frac{1}{2}{({T_d} - T_d^j)^2},
\label{equ:aoi_rpt_area}
\end{align}

\noindent where ${\Delta _j}({t_i}) = {T_u}' + {T_d}'$ equals to the duration of the last round. Then the average AoI $\bar \Delta _j^{RPT}$ is 
\begin{align}
\bar \Delta _j^{RPT} &= \mathop {\lim }\limits_{W \to \infty } \frac{{\sum\nolimits_{w = 1}^W {{A_{(w)}}} }}{{\sum\nolimits_{w = 1}^W {{T_{u(w)}} + {T_{d(w)}}} }} \notag \\
&= \frac{{E\left[ {({T_u}' + {T_d}')({T_u} + T_d^j) + ({T_u} + T_d^j)({T_d} - T_d^j) + \frac{1}{2}{{({T_u} + T_d^j)}^2} + \frac{1}{2}{{({T_d} - T_d^j)}^2}} \right]}}{{E\left[ {{T_u} + {T_d}} \right]}}\notag \\
&= \frac{{{{\left( {E[{T_u}]} \right)}^2} + E[{T_u}]E[T_d^j] + 2E[{T_u}]E[{T_d}] + E[T_d^j]E[{T_d}] + \frac{1}{2}E[{{\left( {{T_u}} \right)}^2}] + \frac{1}{2}E[{{\left( {{T_d}} \right)}^2}]}}{{E[{T_u}] + E[{T_d}]}}.
\label{equ:aoi_rpt}
\end{align}

In the following, we compute each component in (\ref{equ:aoi_rpt}).

\textbf{Computation of $E[{T_u}]$ and $E[{\left( {{T_u}} \right)^2}]$}: Since ${T_u}$ is the duration of the uplink phase and the relay successfully decodes the PNC packet with probability $\alpha $, ${T_u}$ is a geometric random variable with parameter $\alpha $. Thus, we have
\begin{align}
E[{T_u}] = \frac{1}{\alpha },~~~~~~E[{({T_u})^2}] = \frac{{2 - \alpha }}{{{\alpha ^2}}}.
\end{align}

\textbf{Computation of $E[T_d^j]$}: Since $E[T_d^j]$ is the time of packet $C_i^j,j \in \{ A,B\} $ to be recovered (i.e., the PNC packet $C_i^A \oplus C_i^B$ is received) in the downlink phase, $T_d^j$ is also a geometric random variable with parameter $\beta $. Thus, we have 
\begin{align}
E[T_d^j] = \frac{1}{\beta }.
\end{align}

\textbf{Computation of $E[{T_d}]$ and $E[{\left( {{T_d}} \right)^2}]$}: Since ${T_d}$ is the total duration of the downlink phase, by definition, we have 
\begin{align}
{T_d} = \max \{ T_d^A,T_d^B\}.
\end{align}

Since the cumulative distribution function (CDF) of $T_d^j,j \in \{ A,B\} $, is 
\begin{align}
F_d^j(t) = \Pr (T_d^j \le t) = 1 - {(1 - \beta )^t},
\end{align}
\noindent then the CDF of ${T_d}$ is
\begin{align}
{F_d}(t) = \Pr (T_d^{} \le t) = {\left( {1 - {{(1 - \beta )}^t}} \right)^2}.
\end{align}

With the CDF of ${T_d}$, we compute $E[{T_d}]$ and $E[{\left( {{T_d}} \right)^2}]$, respectively 
\begin{align}
E[{T_d}] = \sum\nolimits_{t = 1}^\infty  {\left( {1 - \Pr (T_d^{} \le t - 1)} \right)}  = \sum\nolimits_{t = 1}^\infty  {\left( {1 - {{\left( {1 - {{\left( {1 - \beta } \right)}^{t - 1}}} \right)}^2}} \right)},
\label{equ:aoi_rpt_td_exp1}
\end{align}
\begin{align}
E[{\left( {{T_d}} \right)^2}] = \sum\nolimits_{t = 1}^\infty  {\left( {2t - 1} \right)\left( {1 - \Pr (T_d^{} \le t - 1)} \right)}  = \sum\nolimits_{t = 1}^\infty  {\left( {2t - 1} \right)\left( {1 - {{\left( {1 - {{\left( {1 - \beta } \right)}^{t - 1}}} \right)}^2}} \right)} .
\label{equ:aoi_rpt_td_exp2}
\end{align}

Finally, we substitute all the components into (\ref{equ:aoi_rpt}) to obtain the average AoI $\bar \Delta _j^{RPT}$ 
\begin{align}
\bar \Delta _u^{RPT} = \frac{{\left( \begin{array}{l}
2\alpha  + 4\beta  - \alpha \beta  + \left( {4\alpha \beta  + 2{\alpha ^2}} \right)\sum\nolimits_{t = 1}^\infty  {\left( {1 - {{\left( {1 - {{\left( {1 - \beta } \right)}^{t - 1}}} \right)}^2}} \right)} \\
~~~~~~~~~~~~~~~~~ + {\alpha ^2}\beta \sum\nolimits_{t = 1}^\infty  {\left( {2t - 1} \right)\left( {1 - {{\left( {1 - {{\left( {1 - \beta } \right)}^{t - 1}}} \right)}^2}} \right)} 
\end{array} \right)}}{{2\alpha \beta \left( {1 + \alpha } \right)\sum\nolimits_{t = 1}^\infty  {\left( {1 - {{\left( {1 - {{\left( {1 - \beta } \right)}^{t - 1}}} \right)}^2}} \right)} }}.
\label{equ:aoi_rpt_final}
\end{align}

From (\ref{equ:aoi_rpt_final}), we see that the average AoI of $\bar \Delta _j^{RPT}$ is much complicated than $\bar \Delta _j^{OLTD}$ in (\ref{equ:aoi_oltd_final}). However, when both $\alpha$ and $\beta$ equal to one, it is easy to verify that $\bar \Delta _j^{RPT}=\bar \Delta _j^{OLTD}=3$.

\subsection{Uplink-Lost-Then-Drop (ULTD) Protocol}\label{sec:ARQ2}
Although RPT ensures packet reliability, as will be seen, it leads to high average AoI due to retransmission of old packets in the uplink phase. We now study an uplink-lost-then-drop (ULTD) protocol, combining ARQ and packet drop. Specifically, when the relay fails to decode the PNC packet in the uplink phase, old packets are dropped. The two end users generate and transmit two new packets to the relay in the next time slot immediately. In the downlink phase, ARQ is used to ensure reliable transmission until both users receive the downlink PNC packet. In other words, the uplink transmission of ULTD is the same as that in OLTD, while the downlink transmission is the same as that in RPT.

We compute the average AoI of ULTD, $\bar \Delta _j^{ULTD}, j \in \{ A,B\}$, following the same way as in RPT. Specifically, the area of $\Sigma$ is computed the same as (\ref{equ:aoi_rpt_area}) except that the instantaneous AoI when a successful update occurs becomes ${\Delta _j}({t_i}) = 1 + {T_d}'$ (i.e., time for the uplink phase of a successful update is always one time slot since old packets are dropped in the uplink of ULTD). Then the average AoI $\bar \Delta _j^{ULTD}$ is computed by
\begin{align}
\bar \Delta _j^{ULTD} &= \frac{{E[{T_u}] + E[T_d^j] + 2E[{T_u}]E[{T_d}] + E[T_d^j]E[{T_d}] + \frac{1}{2}E[{{\left( {{T_u}} \right)}^2}] + \frac{1}{2}E[{{\left( {{T_d}} \right)}^2}]}}{{E[{T_u}] + E[{T_d}]}}\notag \\
&= \frac{{\left( \begin{array}{l}
\alpha \beta  + 2{\alpha ^2} + 2\beta  + \left( {4\alpha \beta  + 2{\alpha ^2}} \right)\sum\nolimits_{t = 1}^\infty  {\left( {1 - {{\left( {1 - {{\left( {1 - \beta } \right)}^{t - 1}}} \right)}^2}} \right)} \\
~~~~~~~~~~~~~~~~~~+ {\alpha ^2}\beta \sum\nolimits_{t = 1}^\infty  {\left( {2t - 1} \right)\left( {1 - {{\left( {1 - {{\left( {1 - \beta } \right)}^{t - 1}}} \right)}^2}} \right)} 
\end{array} \right)}}{{2\alpha \beta \left( {1 + \alpha } \right)\sum\nolimits_{t = 1}^\infty  {\left( {1 - {{\left( {1 - {{\left( {1 - \beta } \right)}^{t - 1}}} \right)}^2}} \right)} }}.
\label{equ:aoi_ultd_final}
\end{align}

It is easy to verify that given the same $\alpha$ and $\beta$, $\bar \Delta _j^{ULTD}$ (\ref{equ:aoi_ultd_final}) is smaller than $\bar \Delta _j^{RPT}$ (\ref{equ:aoi_rpt_final}) because $0 < \alpha ,\beta  < 1$. This means that if packet reliability is not a concern (e.g., only information freshness is concerned in many monitoring systems), new packets should be generated and sent immediately when the relay fails to decode the PNC packet in the uplink, in order to achieve high information freshness.

We notice that one can also have a downlink-lost-then-drop (DLTD) protocol. In the uplink, old packets are retransmitted until a PNC packet is successfully decoded at the relay (i.e., ARQ is used in the uplink). In the downlink, the PNC packet is broadcast only once. Whether the downlink PNC packet is decoded or not, the two end users will turn to transmit new packets in the next time slot. In other words, for DLTD, the uplink transmission is the same as that in RPT, while the downlink transmission is the same as that in OLTD. However, it is easy to figure out that the average AoI in DLTD is high because even if the relay can finally decode a PNC packet in the uplink after several retransmissions (e.g., it takes $E[{T_u}]$ time slots on average), old packets will be dropped once the subsequent downlink transmission fails. This leads to an unsuccessful update and the instantaneous AoI continues to increase linearly. Therefore, we omit DLTD in this paper. In the next subsection, we evaluate the theoretical average AoI of OLTD, RPT, and ULTD using RCB.

\subsection{Average AoI Comparison Using RCB}\label{sec:ARQ3}
\begin{figure*}
\centering
\includegraphics[width=0.95\textwidth]{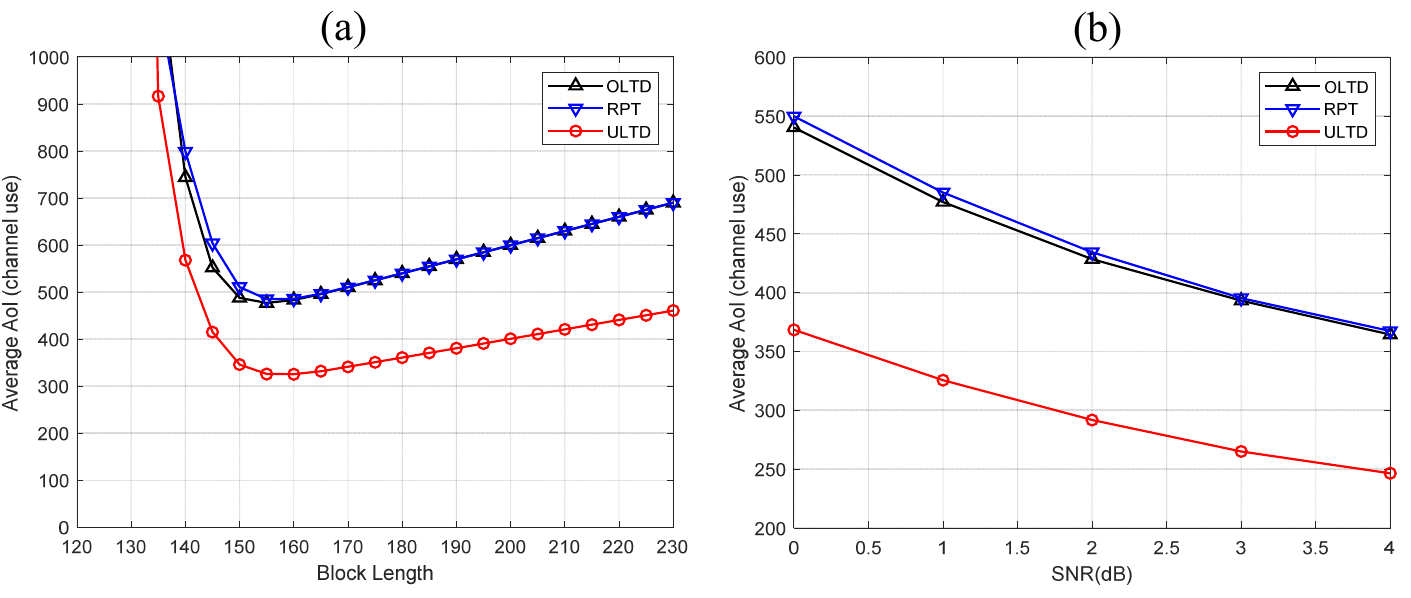}
\caption{(a) The average AoI versus the block length of a coded packet when the SNR is 1dB. (b) The minimum average AoI versus the SNR where the block length is optimized. In both figures, the number of source bits per update packet is $K = 100$.}
\label{fig:result_rcb}
\end{figure*}

In this subsection, we use RCB to estimate the PERs and provide a theoretical average AoI comparison among the three protocols. To estimate $\alpha $ and $\beta $ by RCB, we assume the number of source bits per update packet is  $K = 100$. Fig. \ref{fig:result_rcb}(a) plots the average AoI versus the block length of a coded packet when the SNR is 1dB. The unit of the average AoI is the number of channel use.\footnote{Since the duration of a time slot is proportional to the number of channel use, the average AoI in the unit of channel use is simply the multiplication of the block length and the corresponding average AoI in the unit of time slot.} As depicted in Fig. \ref{fig:result_rcb}(a), for all the three protocols, as the block length increases, the average AoI first decreases and then increases. This is because when the block length is small, the successful packet decoding rates $\alpha $ and $\beta $ are also small, leading to unsuccessful updates most of the time. As the block length increases, the average AoI decreases as more update packets can be received successfully. When the block length is very large such that $\alpha $ and $\beta $ approach to one, the average AoI is also high because the duration of a time slot is large, even if most of the time updates are successful. Therefore, we can see that there is an optimal block length to minimize the average AoI due to the tradeoff between the block length and packet decoding rates. 

Fig. \ref{fig:result_rcb}(b) plots the minimum average AoI versus the SNR. In particular, for each SNR, we optimize the block length to give the minimum average AoI of the three protocols. First, we see that as the SNR increases, the average AoI decreases. Second, as shown in both Fig. \ref{fig:result_rcb}(a) and Fig. \ref{fig:result_rcb}(b), ULTD achieves the minimum average AoI among the three protocols. Specifically, RPT and OLTD almost have the same average AoI, but both are worse than that of ULTD. This indicates that in TWRNs, a pure packet drop protocol (e.g., OLTD, where new packets always have a higher priority) and a pure packet retransmission protocol (e.g., RPT, where old packets always have a higher priority) are not suitable to achieve good information freshness. Instead, combining the ideas from both sides leads to a better average AoI. Such an observation is different from many previous point-to-point information update systems, where dropping old packets leads to a better average AoI performance under the generate-at-will model \cite{aoi_arq}.

The numerical results using RCB are theoretical in nature and only serve to highlight certain points, but do not reflect what actually happens in real wireless communication systems. For example, due to channel fading, the wireless channel is not an AWGN channel in practice. Moreover, the SNR may not be the same during the whole packet duration in a broadband system such as OFDM (e.g., the SNRs are different at different subcarriers). The advantage of ULTD will need to be validated in a real wireless system, as will be presented in the next section.

\section{Experimental Evaluations}\label{sec:exp}
This section presents the experimental evaluation of different transmission protocols in TWRNs, namely OLTD, RPT, and ULTD. Section \ref{sec:exp1} describes the experimental setup on how we conduct real experiments using software-defined radios. Section \ref{sec:exp2} details the average AoI performance of different protocols. Furthermore, we discuss the difference between AoI and conventional system metrics using the experimental results, such as throughput and delay.

\begin{figure}
\centering
\includegraphics[width=0.65\textwidth]{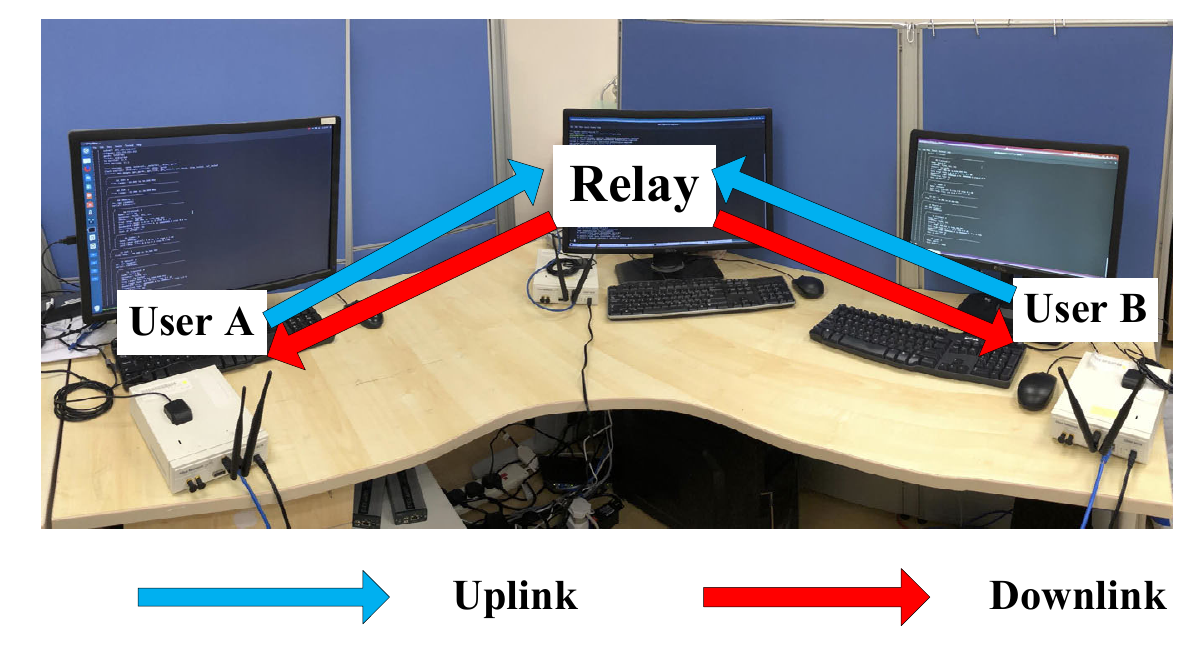}
\caption{An indoor information update TWRN system built upon the OFDM-based real-time PNC system \cite{liew2015primer}.}
\label{fig:exp_setup}
\end{figure}

\subsection{Experiment Setup}\label{sec:exp1}
For experimentation, our system adopts the USRP hardware (USRP N210 with SBX daughterboards) and the GNU Radio software with the UHD hardware driver. We consider an indoor information update TWRN system as shown in Fig. \ref{fig:exp_setup}. Our system is built upon the OFDM-based real-time PNC system \cite{liew2015primer}. We examine the average AoI performance by employing trace-driven simulations using PHY-layer decoding results. Specifically, we first obtain the decoding outcomes in a series of time slots. The PNC decoding outcomes at the relay in the uplink and the broadcast PNC packet decoding outcomes at the end users in the downlink are gathered. Then, we generate traces based on the PHY-layer decoding outcomes to drive our AoI simulations with different protocols.

To gather the PHY-layer decoding results, our experiments are carried out at 2.185 GHz center frequency with 5 MHz bandwidth. In the uplink phase, the relay sends a beacon frame to trigger the simultaneous transmission of a series of update packets from the two end users (e.g., 2000 update packets in our experiments). We control the transmit powers of the two end users so that they almost have the same SNR at the relay in the uplink. The received SNR is varied from 7.5dB to 10dB. Notice that the received SNR of an end user’s update packets could be slightly different due to channel fading. The SNR here is the average SNR of the 2000 update packets.

Both end users adopt the BPSK modulation and the standard rate-1/2 $[133, 171]_8$ convolutional code used in the 802.11 standard \cite{dot11std13}. Since the received SNR is an average value, we choose the fixed-rate convolutional code in our experiments (e.g., optimizing the block length based on instantaneous SNR values to minimize the average AoI may not be possible or practical). The number of source bits per update packet is $K=100$ since packets containing update information from IoT devices are typically short (e.g., tens of bytes) in practice. For each SNR, we collect the PNC decoding outcomes of the 2000 superimposed packets at the relay. 

Similarly, in the downlink phase, we adjust the positions of the end users so that the relay has the same received SNR at the two end users in the downlink. The relay sends 2000 packets to the end users at each SNR, also ranging from 7.5dB to 10dB. The two end users decode the downlink packets and save the decoding outcomes. 

\subsection{Experiment Results}\label{sec:exp2}
\begin{figure}
\centering
\includegraphics[width=0.62\textwidth]{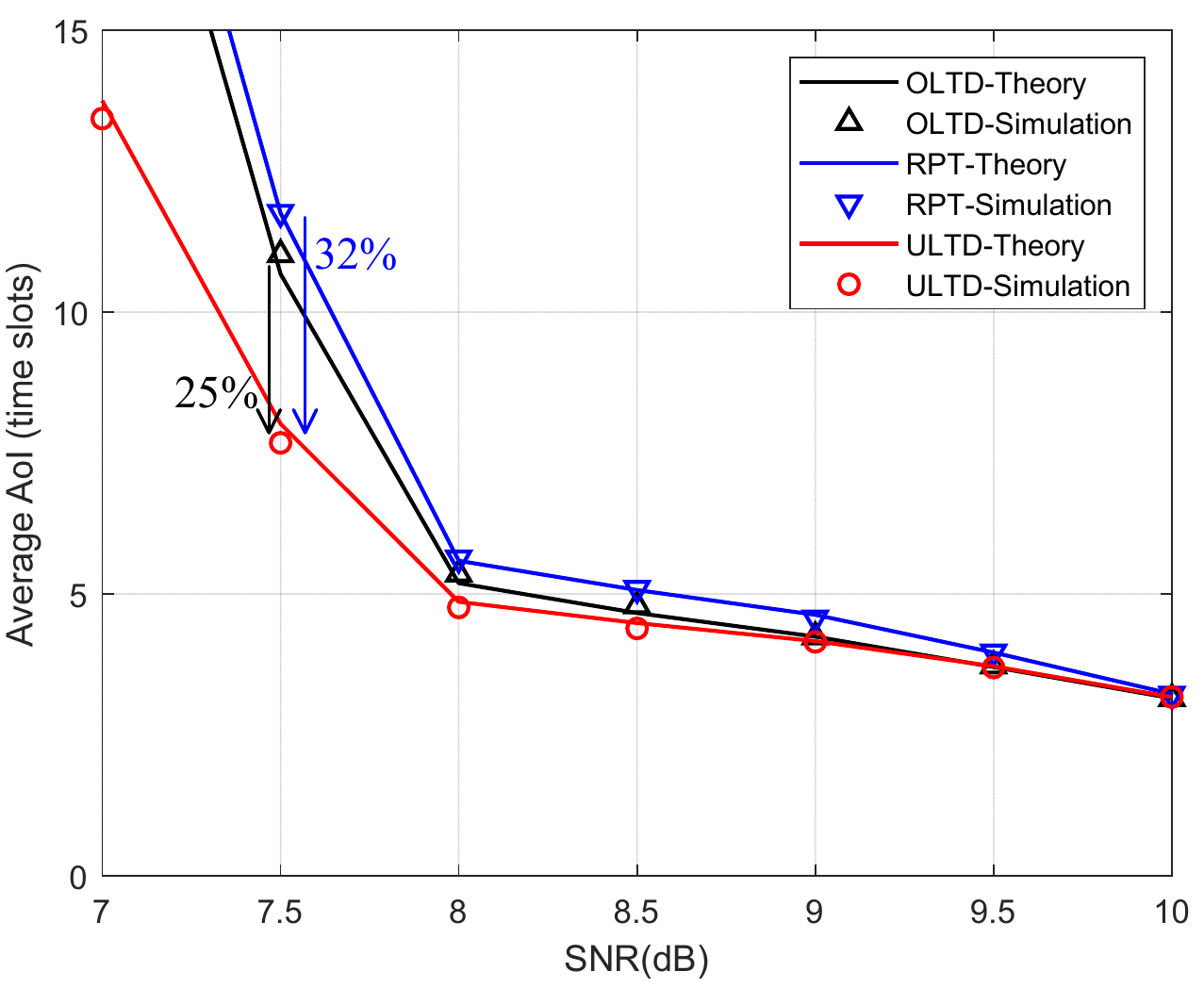}
\caption{Experimental results of the average AoI versus the SNR on the software-defined radio under the three protocols considered in this paper. ULTD has a lower average AoI than that of OLTD and RPT.}
\label{fig:result_exp}
\end{figure}

Fig. \ref{fig:result_exp} plots the average AoI of OLDT, RPT, and ULTD. Specifically, for the theoretical results, we first collect the successful packet decoding rates in the uplink and the downlink ($\alpha$ and $\beta$), and then calculate the average AoI based on the average AoI formulas obtained in previous sections (i.e., (\ref{equ:aoi_oltd_final}), (\ref{equ:aoi_rpt_final}), and (\ref{equ:aoi_ultd_final})). For the simulations results, we collect the packet decoding outcome in each time slot, from which we compute the instantaneous AoI, followed by the average AoI. Note here that the unit of the average AoI is the number of time slots (e.g., the packet duration remains a constant).

Overall, we see from Fig. \ref{fig:result_exp} that the simulation results corroborate with the theoretical results, indicating the correctness of the average AoI derived in previous sections. In addition, as in Fig. \ref{fig:result_rcb} in which RCB is used, ULTD outperforms both RPT and OLTD. In our experiments, the performance improvement of ULTD is more significant when the SNR is low. Specifically, ULTD reduces the average AoI by 25\% and 32\% at the SNR of 7.5dB compared with OLTD and RPT, respectively. As the SNR increases, all the three protocols tend to have the same average AoI since the PER approaches zero, including both uplink and downlink transmissions. 

Our experiments on software define radio indicate that in a practical setting, ULTD protocol is a preferable protocol to achieve high information freshness in a TWRN operated with PNC. More specifically, in such a two-hop TWRN, when old packets get corrupted, dropping the old packets should be adopted in the first hop (i.e., in the uplink). If we drop the old packets in the first hop, since a newly sent packet contains fresher information, ULTD will always have a smaller instantaneous AoI once the packet is finally received successfully at the destination, compared with RPT. This is the reason why ULTD outperforms RPT. Although the instantaneous AoI upon a successful update is always two time slots for OLTD (which is the lowest possible instantaneous AoI), OLTD always generates new packets once the second hop fails. It takes time for the relay to decode a new PNC packet and then go to the second hop again, and thus the time interval of receiving two consecutive update packets could be very large in OLTD (leading to high instantaneous AoI). As a result, the average AoI of OLTD is much larger than that of ULTD, indicating that packet retransmission should be used in the second hop, as ULTD does.

\begin{figure}
\centering
\includegraphics[width=1\textwidth]{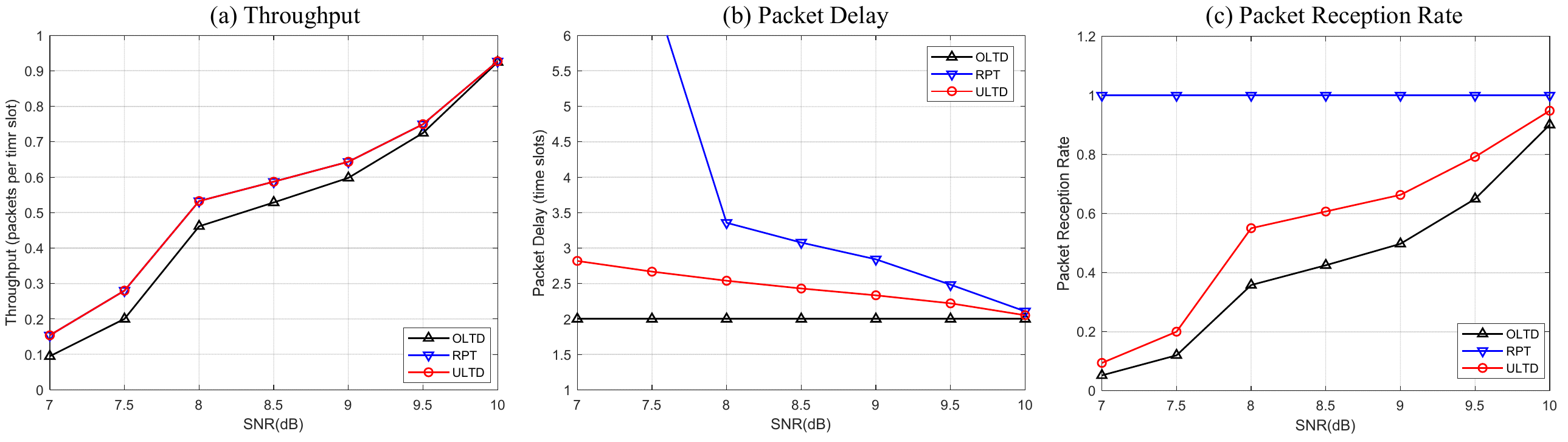}
\caption{Experimental results of the (a) average AoI, (b) packet delay, and (c) packet reception rate versus the SNR on the software defined radio under the three protocols considered in this paper. }
\label{fig:other_metric}
\end{figure}

 \emph{\textbf{Discussion: Differences between AoI and Conventional Metrics}} $-$ We now give a discussion on the differences between the new AoI metric and conventional metrics. Let us first consider throughput as the system performance metric in Fig. \ref{fig:other_metric}(a). It is easy to figure out that since OLTD drops old packets once old packets are corrupted, OLTD has the lowest system throughput among the three schemes, in terms of the number of received update packets per time slot, as shown in Fig. \ref{fig:other_metric}(a). Furthermore, RPT and ULTD have the same total throughput, e.g., the system throughputs of both protocols are $2/\left( {E[{T_u}] + E[T_d^{}]} \right)$ packets per time slot. However, Fig. \ref{fig:result_exp} shows that ULTD provides a much better average AoI than RPT does. This shows that with the same throughput, dropping old packets can increase information freshness in ULTD.

As far as delay is concerned, Fig. \ref{fig:other_metric}(b) compares the packet delay for the successfully decoded packets of the three protocols. The packet delay is defined as the duration from the time when the packet is sent by one user for the first time to the time when the packet is finally received by the other user. Unlike average AoI, OLTD has the smallest packet delay among the three protocols. Specifically, the packet delay for all the decoded packets is always two time slots in OLTD (i.e., the meaning of “2” in the average AoI (\ref{equ:aoi_ultd_final})). This also means that if an information update system wants to minimize the lowest instantaneous AoI, OLTD should be adopted, e.g., for each successful update, the received packet always contains the most up-to-date information. In addition, the average packet delays for RPT and ULTD are $E[{T_u}] + E[T_d^j]$ and $1 + E[T_d^j]$, respectively, i.e., ULTD has a shorter packet delay than that of RPT, as shown in Fig. \ref{fig:other_metric}. 

Fig. \ref{fig:other_metric}(c) plots the packet reception rates of the three protocols. The packet reception rate is defined as the number of packets received by the destination divided by the total number of packets generated at the source under the generate-at-will model. In general, RPT receives all the update packets by ARQ, while OLTD has the lowest packet reception rate since it drops old packets once an uplink or a downlink transmission fails. By contrast, the packet reception rate of ULTD sits between RPT and OLTD. Since ULTD provides the lowest average AoI and a moderate packet reception rate, it is a viable solution to information update systems with packet reception rate requirements. For example, in vehicular networks, vehicles exchange their basic safety message (BSM) to each other via a road side unit (RSU) as a relay. The BSM update packets sent by vehicles contain their latest status information, such as position. Not only does a vehicle require high information freshness to acquire the current positions of other vehicles, but also needs a certain number of received packets to infer the trajectories of other vehicles to better help improve road safety. In other words, it is critical to receive BSM update packets with high information freshness and high packet reception rates, and ULTD can satisfy both requirements.

\section{Conclusions}\label{sec:conclusions}
In this paper, we have studied the average AoI of a PNC-enabled TWRN with and without ARQ. In particular, we put forth an uplink-lost-then-drop (ULTD) protocol that combines packet drop and ARQ to achieve high information freshness.

PNC is a key technique that turns the superimposed wireless signals into network-coded messages, thereby reducing the communication delay of TWRNs. With a shorter delay, end users can send update packets to each other more frequently, thereby improving information freshness. However, when update packets get corrupted in either hop in a TWRN, one needs to decide the packet to be dropped or to be retransmitted, because new packets always contain recent information but they may require more time to be delivered. Dealing with the corrupted packets to achieve low average AoI in each hop of a TWRN is the main focus of this paper. 

We find that neither a non-ARQ scheme nor a pure ARQ scheme achieves good average AoI performance. Theoretical analysis shows that, unlike single-hop networks, a non-ARQ scheme where old packets are dropped when they get corrupted (i.e., OLTD) improves the average AoI little over a classical ARQ scheme with no packet lost (i.e., RPT). Hence, we combine packet drop and ARQ: in ULTD, corrupted packets are dropped in the uplink, but they will be retransmitted until successful reception in the downlink. We believe the insight of ULTD applies generally to other two-hop networks. Experiments on software-defined radio indicate that the average AoI of ULTD is significantly lower than that of OLTD and RPT. Moreover, our experimental results show the key differences between AoI and conventional performance metrics, such as throughput and delay. 

\appendices
\section{Computation of Soft Information in the XOR-CD Decoder}\label{sec:PNC}
This appendix details the soft information computation, in terms of Log-Likelihood Ratio (LLR), in the XOR-CD decoder. In the uplink, packets $C^A$ from user A and $C^B$ from user B are channel-encoded into $V^A$ and $V^B$, respectively. Let $V^j = (v^j[1],...,v^j[z],...), j\in\{A,B\}$ denote the PHY-layer codeword of user $j$, where $v^j[z]$ is the $z$-th encoded bit. $V^j$ is then BPSK-modulated into $X^j = (x^j[1],...,x^j[z],...)$, where $x^j[z] = 1 - 2v^j[z]$ is the $z$-th BPSK symbol. 

We assume an OFDM system where multipath fading can be dealt with by cyclic prefix (CP). The $z$-th received sample in the frequency domain at the receiver can be written as
\begin{align}
y[z] = h^{A}[z]x^{A}[z] + h^B[z]x^B[z] + w[z],
\label{equ:received_sample}
\end{align}

\noindent where $h^{A}[z]$ and $h^{B }[z]$  are the channel gains from users A and B at the relay, respectively, and $w[n]$ is the Gaussian noise.

Recall that XOR-CD modifies the computation of the LLRs of bits, without modifying the Viterbi decoding algorithm that makes use of the LLRs (see Section \ref{sec:preliminaries1}). The LLR computation is based on a \emph{reduced-constellation demodulation principle} \cite{liew2015primer}. If we drop the index $z$, the LLR of the XOR bit ${v^A} \oplus {v^B}$ is given by 
\begin{align}
LLR({v^A} \oplus {v^B}) &= \log \frac{{\Pr ({v^A} \oplus {v^B} = 0|y)}}{{\Pr ({v^A} \oplus {v^B} = 1|y)}},
\notag\\
&\propto \log \frac{{\Pr ({x^A} = 1,{x^B} = 1|y) + \Pr ({x^A} =  - 1,{x^B} =  - 1|y)}}{{\Pr ({x^A} = 1,{x^B} =  - 1|y) + \Pr ({x^A} =  - 1,{x^B} = 1|y)}},
\label{equ:llr} \\
&\propto \max \{  - |y - h^A - h^B{|^2}, - |y + h^A + h_i^B{|^2}\} \notag \\
&~~~~~~~~~~~~~~~~~- \max \{  - |y - h^A + h^B{|^2}, - |y + h^A - h^B{|^2}\} \label{equ:llr_max} \\
&\approx \min \{ |y - {h^A} + {h^B}{|},|y + {h^A} - {h^B}{|}\}, \notag \\
&~~~~~~~~~~~~~~~~~ - \min \{ |y - {h^A} - {h^B}{|},|y + {h^A} + {h^B}{|}\}. \label{equ:llr_min}
\end{align}

Equation (\ref{equ:llr}) includes the computations of the probabilities of the four constellation points with ${x^A} =  \pm 1$ and ${x^B} =  \pm 1$. From (\ref{equ:llr}) to (\ref{equ:llr_min}), we use the \emph{log-max} approximation for further simplification. As a result, (\ref{equ:llr_min}) can be interpreted as selecting the two most likely constellation points from the four constellation points induced by the concurrent transmissions of users A and B. The two points are chosen based on the minimum Euclidean distance to represent possible the two possible XOR bits ${v^A} \oplus {v^B} = 0$ and ${v^A} \oplus {v^B} = 1$. After that, LLRs of $V^A \oplus V^B = (v^A[1] \oplus v^B[1],...,v^A[z] \oplus v^B[z],...)$ are fed into the Viterbi decoder to decode $C^A \oplus C^B$.

\bibliographystyle{IEEEtran}
\bibliography{aoi_twrn}

\begin{thebibliography}{10}
\providecommand{\url}[1]{#1}
\csname url@samestyle\endcsname
\providecommand{\newblock}{\relax}
\providecommand{\bibinfo}[2]{#2}
\providecommand{\BIBentrySTDinterwordspacing}{\spaceskip=0pt\relax}
\providecommand{\BIBentryALTinterwordstretchfactor}{4}
\providecommand{\BIBentryALTinterwordspacing}{\spaceskip=\fontdimen2\font plus
\BIBentryALTinterwordstretchfactor\fontdimen3\font minus
  \fontdimen4\font\relax}
\providecommand{\BIBforeignlanguage}[2]{{%
\expandafter\ifx\csname l@#1\endcsname\relax
\typeout{** WARNING: IEEEtran.bst: No hyphenation pattern has been}%
\typeout{** loaded for the language `#1'. Using the pattern for}%
\typeout{** the default language instead.}%
\else
\language=\csname l@#1\endcsname
\fi
#2}}
\providecommand{\BIBdecl}{\relax}
\BIBdecl

\bibitem{aoi_mono}
Y.~{Sun}, I.~{Kadota}, R.~{Talak}, E.~{Modiano}, and R.~{Srikant}, \emph{Age of
  Information: A New Metric for Information Freshness}.\hskip 1em plus 0.5em
  minus 0.4em\relax Morgan \& Claypool, 2019.

\bibitem{AoIMagazine}
M.~A. {Abd-Elmagid}, N.~{Pappas}, and H.~S. {Dhillon}, ``On the role of age of
  information in the internet of things,'' \emph{IEEE Communications Magazine},
  vol.~57, no.~12, pp. 72--77, Dec. 2019.

\bibitem{AoI}
A.~Kosta, N.~Pappas, and V.~Angelakis, ``Age of information: A new concept,
  metric, and tool,'' \emph{Foundations and Trends® in Networking}, vol.~12,
  no.~3, pp. 162--259, 2017.

\bibitem{aoi_first}
S.~{Kaul}, M.~{Gruteser}, V.~{Rai}, and J.~{Kenney}, ``Minimizing age of
  information in vehicular networks,'' in \emph{Proc. IEEE SECON}, 2011, pp.
  350--358.

\bibitem{IIOT}
E.~{Sisinni}, A.~{Saifullah}, S.~{Han}, U.~{Jennehag}, and M.~{Gidlund},
  ``Industrial internet of things: Challenges, opportunities, and directions,''
  \emph{IEEE Transactions on Industrial Informatics}, vol.~14, no.~11, pp.
  4724--4734, 2018.

\bibitem{PNC06}
S.~Zhang, S.~C. Liew, and P.~P. Lam, ``Hot topic: physical-layer network
  coding,'' in \emph{Proc. ACM MOBICOM}, 2006, pp. 358--365.

\bibitem{hybridmac}
M.~Zhang, G.~G. M.~N. Ali, P.~H.~J. Chong, B.-C. Seet, and A.~Kumar, ``A novel
  hybrid mac protocol for basic safety message broadcasting in vehicular
  networks,'' \emph{IEEE Transactions on Intelligent Transportation Systems},
  vol.~21, no.~10, pp. 4269--4282, Oct. 2020.

\bibitem{liew2015primer}
S.~C. Liew, L.~Lu, and S.~Zhang, \emph{A Primer on Physical-Layer Network
  Coding}.\hskip 1em plus 0.5em minus 0.4em\relax Morgan \& Claypool Publishers
  Synthesis Lectures on Communication Networks, 2015.

\bibitem{gallager}
R.~G. Gallager, \emph{Information Theory and Reliable Communication}.\hskip 1em
  plus 0.5em minus 0.4em\relax New York, NY, USA: John Wiley \& Sons, Inc.,
  1968.

\bibitem{tse2005fundamentals}
D.~Tse and P.~Viswanath, \emph{Fundamentals of Wireless Communication}.\hskip
  1em plus 0.5em minus 0.4em\relax Cambridge university press, 2005.

\bibitem{aoi_arq}
M.~Xie, Q.~Wang, J.~Gong, and X.~Ma, ``Age and energy analysis for {LDPC} coded
  status update with and without {ARQ},'' \emph{IEEE Internet of Things
  Journal}, vol.~7, no.~10, pp. 10\,388--10\,400, Oct. 2020.

\bibitem{AoI_single_server}
Y.~{Inoue}, H.~{Masuyama}, T.~{Takine}, and T.~{Tanaka}, ``A general formula
  for the stationary distribution of the age of information and its application
  to single-server queues,'' \emph{IEEE Transactions on Information Theory},
  vol.~65, no.~12, pp. 8305--8324, Dec. 2019.

\bibitem{queue1}
S.~{Kaul}, R.~{Yates}, and M.~{Gruteser}, ``Real-time status: How often should
  one update?'' in \emph{2012 Proceedings IEEE INFOCOM}, 2012, pp. 2731--2735.

\bibitem{AoIqueue}
A.~M. {Bedewy}, Y.~{Sun}, and N.~B. {Shroff}, ``Minimizing the age of
  information through queues,'' \emph{IEEE Transactions on Information Theory},
  vol.~65, no.~8, pp. 5215--5232, Aug 2019.

\bibitem{AoIscheduling4}
I.~{Kadota}, A.~{Sinha}, E.~{Uysal-Biyikoglu}, R.~{Singh}, and E.~{Modiano},
  ``Scheduling policies for minimizing age of information in broadcast wireless
  networks,'' \emph{IEEE/ACM Transactions on Networking}, vol.~26, no.~6, pp.
  2637--2650, Dec 2018.

\bibitem{AoIscheduling3}
C.~{Joo} and A.~{Eryilmaz}, ``Wireless scheduling for information freshness and
  synchrony: Drift-based design and heavy-traffic analysis,'' \emph{IEEE/ACM
  Transactions on Networking}, vol.~26, no.~6, pp. 2556--2568, Dec 2018.

\bibitem{AoIscheduling2}
I.~{Kadota}, A.~{Sinha}, and E.~{Modiano}, ``Scheduling algorithms for
  optimizing age of information in wireless networks with throughput
  constraints,'' \emph{IEEE/ACM Transactions on Networking}, vol.~27, no.~4,
  pp. 1359--1372, Aug 2019.

\bibitem{AoIscheduling1}
N.~Lu, B.~Ji, and B.~Li, ``Age-based scheduling: Improving data freshness for
  wireless real-time traffic,'' in \emph{Proc. ACM MOBICOM}, New York, NY, USA,
  2018, pp. 191--200.

\bibitem{peak_aoi}
J.~P. Champati, R.~R. Avula, T.~J. Oechtering, and J.~Gross, ``Minimum
  achievable peak age of information under service preemptions and request
  delay,'' \emph{IEEE Journal on Selected Areas in Communications}, vol.~39,
  no.~5, pp. 1365--1379, May 2021.

\bibitem{aoi_tdma_fdma}
H.~Pan and S.~C. Liew, ``Information update: {TDMA} or {FDMA}?'' \emph{IEEE
  Wireless Communications Letters}, vol.~9, no.~6, pp. 856--860, June 2020.

\bibitem{aoi_harq}
E.~T. {Ceran}, D.~{Gündüz}, and A.~{György}, ``Average age of information
  with hybrid {ARQ} under a resource constraint,'' \emph{IEEE Transactions on
  Wireless Communications}, vol.~18, no.~3, pp. 1900--1913, March 2019.

\bibitem{chenhe_aoi}
Y.~{Gu}, H.~{Chen}, Y.~{Zhou}, Y.~{Li}, and B.~{Vucetic}, ``Timely status
  update in internet of things monitoring systems: An age-energy tradeoff,''
  \emph{IEEE Internet of Things Journal}, vol.~6, no.~3, pp. 5324--5335, June
  2019.

\bibitem{steam_code}
H.~Pan, S.~C. Liew, J.~Liang, V.~C.~M. Leung, and J.~Li, ``Coding of
  multi-source information streams with age of information requirements,''
  \emph{IEEE Journal on Selected Areas in Communications}, vol.~39, no.~5, pp.
  1427--1440, May 2021.

\bibitem{aoi_multicast}
M.~Xie, J.~Gong, X.~Jia, and X.~Ma, ``Age and energy tradeoff for multicast
  networks with short packet transmissions,'' \emph{IEEE Transactions on
  Communications}, pp. 1--1, 2021.

\bibitem{control_drop}
V.~Kavitha, E.~Altman, and I.~Saha, ``Controlling packet drops to improve
  freshness of information,'' \emph{Technical report, \textnormal{available
  at:} \textnormal{\url{https://arxiv.org/abs/1807.09325}}}.

\bibitem{aoi_adaptive_coding}
S.~{Feng} and J.~{Yang}, ``Adaptive coding for information freshness in a
  two-user broadcast erasure channel,'' in \emph{IEEE Global Communications
  Conference (GLOBECOM)}, 2019, pp. 1--6.

\bibitem{aoi_multihop}
A.~M. {Bedewy}, Y.~{Sun}, and N.~B. {Shroff}, ``The age of information in
  multihop networks,'' \emph{IEEE/ACM Transactions on Networking}, vol.~27,
  no.~3, pp. 1248--1257, June 2019.

\bibitem{BoostingorHindering}
J.~Lou, X.~Yuan, S.~Kompella, and N.-F. Tzeng, ``Boosting or hindering: {AoI}
  and throughput interrelation in routing-aware multi-hop wireless networks,''
  \emph{IEEE/ACM Transactions on Networking}, vol.~29, no.~3, pp. 1008--1021,
  June 2021.

\bibitem{mangang_relay1}
M.~Xie, J.~Gong, and X.~Ma, ``Age and energy tradeoff for short packet based
  two-hop decode-and-forward relaying networks,'' in \emph{IEEE Wireless
  Communications and Networking Conference (WCNC)}, 2021, pp. 1--6.

\bibitem{mangang_relay2}
------, ``Age-energy tradeoff in dual-hop status update systems with the m-th
  best relay selection,'' in \emph{IEEE 93rd Vehicular Technology Conference
  (VTC2021-Spring)}, 2021, pp. 1--5.

\bibitem{aoi_noma_tii}
H.~Pan, J.~Liang, S.~C. Liew, V.~C.~M. Leung, and J.~Li, ``Timely information
  update with nonorthogonal multiple access,'' \emph{IEEE Transactions on
  Industrial Informatics}, vol.~17, no.~6, pp. 4096--4106, June 2021.

\bibitem{aoi_af}
B.~Li, H.~Chen, N.~Pappas, and Y.~Li, ``Optimizing information freshness in
  two-way relay networks,'' in \emph{IEEE/CIC International Conference on
  Communications in China (ICCC)}, 2020, pp. 893--898.

\bibitem{short_pnc}
S.~Salamat~Ullah, S.~C. Liew, G.~Liva, and T.~Wang, ``Short-packet
  physical-layer network coding,'' \emph{IEEE Transactions on Communications},
  vol.~68, no.~2, pp. 737--751, Feb. 2020.

\bibitem{dot11std13}
{IEEE802.11ac-2013}, ``Wireless {LAN} medium access control ({MAC}) and
  physical layer ({PHY}) specifications amendment 4: Enhancements for very high
  throughput for operation in bands below 6 {GH}z.''

\end{thebibliography}
\end{document}